\documentclass[aps,prl,preprint,amsmath,amssymb,showpacs,superscriptaddress]{revtex4-2}

\usepackage{graphicx}% Include figure files
\usepackage{dcolumn}% Align table columns on decimal point
\usepackage{bm}
\usepackage{graphicx}
\usepackage{overpic}
\usepackage{longtable}
\usepackage{epsfig}
\usepackage{epsf}
\usepackage{booktabs}
\usepackage{colortbl}
\usepackage{booktabs}
\usepackage{amssymb}
\usepackage{amsmath}
\usepackage{amsthm}
\usepackage{multirow}
\usepackage{verbatim}
\usepackage[T1]{fontenc}
\usepackage{amsfonts}
\usepackage{scalerel}
\usepackage{cases}
\usepackage{pgfplots}
\usepackage{float}
\usepackage{lipsum}
\usepackage{gensymb}
\usepackage[colorlinks=true,linkcolor=blue,citecolor=blue]{hyperref}

\newcommand{\myfont}{\fontfamily{phv}\selectfont}
\newcommand{\figfont}{\myfont \textsf \large}

\makeatletter
\newenvironment{figurehere}{\def\@captype{figure}}

\begin{document}

\title{99.92\%-Fidelity CNOT Gates in Solids by Filtering Time-dependent and Quantum Noises}

\author{Tianyu Xie}
\thanks{T. Xie and Z. Zhao contributed equally to this work.}
\affiliation{CAS Key Laboratory of Microscale Magnetic Resonance and School of Physical Sciences, University of Science and Technology of China, Hefei 230026, China}
\affiliation{CAS Center for Excellence in Quantum Information and Quantum Physics, University of Science and Technology of China, Hefei 230026, China}

\author{Zhiyuan Zhao}
\thanks{T. Xie and Z. Zhao contributed equally to this work.}
\affiliation{CAS Key Laboratory of Microscale Magnetic Resonance and School of Physical Sciences, University of Science and Technology of China, Hefei 230026, China}
\affiliation{CAS Center for Excellence in Quantum Information and Quantum Physics, University of Science and Technology of China, Hefei 230026, China}

\author{Shaoyi Xu}
\affiliation{CAS Key Laboratory of Microscale Magnetic Resonance and School of Physical Sciences, University of Science and Technology of China, Hefei 230026, China}
\affiliation{CAS Center for Excellence in Quantum Information and Quantum Physics, University of Science and Technology of China, Hefei 230026, China}

\author{Xi Kong}
\affiliation{National Laboratory of Solid State Microstructures and Department of Physics, Nanjing University, Nanjing 210093, China}

\author{Zhiping Yang}
\affiliation{CAS Key Laboratory of Microscale Magnetic Resonance and School of Physical Sciences, University of Science and Technology of China, Hefei 230026, China}
\affiliation{CAS Center for Excellence in Quantum Information and Quantum Physics, University of Science and Technology of China, Hefei 230026, China}

\author{Mengqi Wang}
\affiliation{CAS Key Laboratory of Microscale Magnetic Resonance and School of Physical Sciences, University of Science and Technology of China, Hefei 230026, China}
\affiliation{CAS Center for Excellence in Quantum Information and Quantum Physics, University of Science and Technology of China, Hefei 230026, China}

\author{Ya Wang}
\affiliation{CAS Key Laboratory of Microscale Magnetic Resonance and School of Physical Sciences, University of Science and Technology of China, Hefei 230026, China}
\affiliation{CAS Center for Excellence in Quantum Information and Quantum Physics, University of Science and Technology of China, Hefei 230026, China}
\affiliation{Hefei National Laboratory, University of Science and Technology of China, Hefei 230088, China}

\author{Fazhan Shi}
\email{fzshi@ustc.edu.cn}
\affiliation{CAS Key Laboratory of Microscale Magnetic Resonance and School of Physical Sciences, University of Science and Technology of China, Hefei 230026, China}
\affiliation{CAS Center for Excellence in Quantum Information and Quantum Physics, University of Science and Technology of China, Hefei 230026, China}
\affiliation{Hefei National Laboratory, University of Science and Technology of China, Hefei 230088, China}
\affiliation{School of Biomedical Engineering and Suzhou Institute for Advanced Research, University of Science and Technology of China, Suzhou 215123, China}

\author{Jiangfeng Du}
\email{djf@ustc.edu.cn}
\affiliation{CAS Key Laboratory of Microscale Magnetic Resonance and School of Physical Sciences, University of Science and Technology of China, Hefei 230026, China}
\affiliation{CAS Center for Excellence in Quantum Information and Quantum Physics, University of Science and Technology of China, Hefei 230026, China}
\affiliation{Hefei National Laboratory, University of Science and Technology of China, Hefei 230088, China}

\clearpage

\begin{abstract}
Inevitable interactions with the reservoir largely degrade the performance of non-local gates, which hinders practical quantum computation from coming into existence. Here we experimentally demonstrate a 99.920(7)\%-fidelity controlled-NOT gate by suppressing the complicated noise in a solid-state spin system at room temperature. We found that the fidelity limited at 99\% in previous works results from only considering static noise, and thus, in this work, time-dependent noise and quantum noise are also included. All noises are dynamically corrected by an exquisitely designed shaped pulse, giving the resulting error below $10^{-4}$. The residual gate error is mainly originated from the longitudinal relaxation and the waveform distortion that can both be further reduced technically. Our noise-resistant method is universal, and will benefit other solid-state spin systems.
\end{abstract}

\maketitle

High-fidelity non-local gates play a crucial role in quantum information processing, in particular, fault-tolerant quantum computation~\cite{nielsen2000quantum}. However, inevitable interactions with the reservoir dramatically degrade the performance of the desired gates, especially for solid-state systems. After decades of efforts, quantum systems like superconducting circuits~\cite{barends2014superconducting, arute2019quantum, wu2021strong}, trapped ions~\cite{hughes2020benchmarking, pino2021demonstration, clark2021high}, solid-state defects~\cite{rong2015experimental, bradley2019ten, mkadzik2022precision}, and quantum dots~\cite{noiri2022fast, xue2022quantum}, have demonstrated the non-local gates with fidelities above the surface-code threshold (about 99\%) for fault tolerance~\cite{wang2011surface}, but practical large-scale quantum computation demands a fidelity of at least 99.9\%~\cite{fowler2012surface}.

The gate errors on logical qubits can be rendered arbitrarily small by quantum error correction (QEC) provided that the errors on physical qubits are below a certain threshold~\cite{nielsen2000quantum, wang2011surface, fowler2012surface}. Similarly, the physical errors can be further corrected dynamically among multiple primitive gates once the non-Markovian nature embraced by the noise is reasonably harnessed~\cite{khodjasteh2009dynamically, khodjasteh2010arbitrarily}. Therefore, by combining these two levels of error correction, the fidelity threshold on primitive gates could be substantially lowered, thus providing a feasible approach to practical quantum computation supported by real-world imperfect devices. However, it is experimentally hard to achieve higher-fidelity gates after error correction owing to much extra circuit complexity incurred by the correction process. Recently, universal gates on logical qubits have been demonstrated but still perform much worse than those on physical qubits~\cite{postler2022demonstration}. The same situation appears when the method of dynamical error correction (DEC) is applied for non-local gates, because the noise characteristics for cancelling the errors in between control pulses are not precisely grasped~\cite{rong2015experimental, waldherr2014quantum, dolde2014high}, and more pulses and longer gate times may result in larger errors.

Over the past decades, the high-fidelity operations enhanced by DEC are mainly single-qubit gates~\cite{rong2015experimental, soare2014experimental, cerfontaine2020closed, werninghaus2021leakage}, where the special case is the Identity operation spanning a long duration through dynamical decoupling~\cite{du2009preserving, biercuk2009optimized, de2010universal}. Although there are many theoretical schemes to realize DEC-improved two-qubit gates~\cite{montangero2007robust, allen2017optimal, chou2015optimal, huang2019high}, experimental realization remains intractable due to idealized system description and noise modelling. In this work, the reservoir noise is precisely measured, and nearly two orders of magnitude suppressed via DEC in a solid-state spin system, namely the nitrogen-vacancy (NV) center in diamond. The fidelity of the CNOT gate realized here is improved from 99.52(2)\% (primitive) to 99.920(7)\%, both estimated by the method of randomized benchmarking~\cite{magesan2011scalable, magesan2012efficient}.

\begin{figure}
\centering
\begin{overpic}[width=1.0\columnwidth]{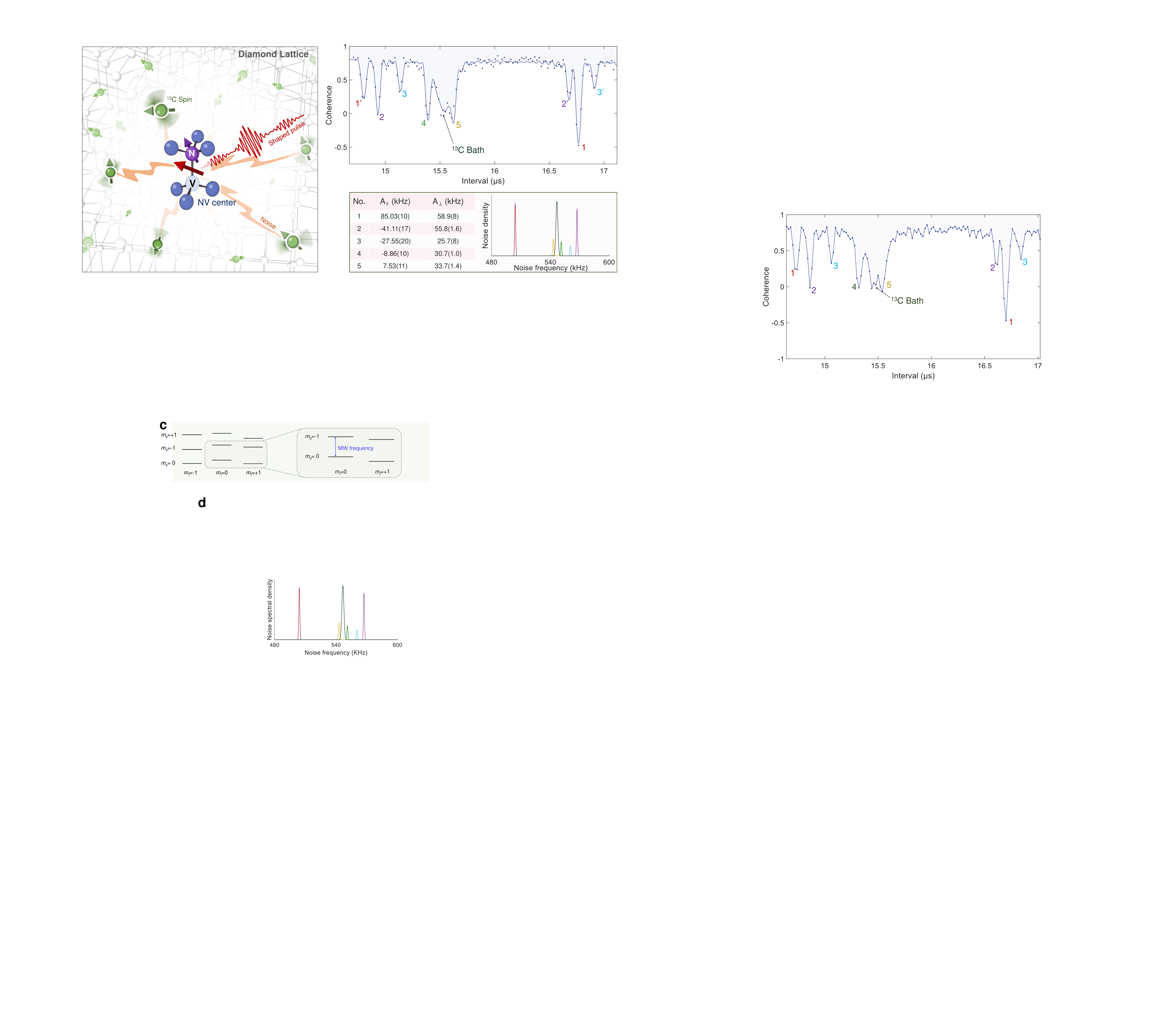}
\renewcommand{\familydefault}{\sfdefault}
	\put (0.0, 42.5) {\figfont{(a)}}
	\put (46.0, 42.5) {\figfont{(b)}}
	\put (48, 15.8) {\figfont{(c)}}
\end{overpic}
\caption{Characterization of the solid-state spin system and its reservoir. (a) Schematic of a single NV center in diamond lattice. The coherence of the NV electron spin is severely disturbed by the surrounding $^{13}$C nuclear spins. The proximal five $^{13}$C nuclear spins are strongly coupled to the NV center, and thus taken as quantum noise, as measured and displayed in (b) and (c). The other $^{13}$C spins are effectively considered as classical noise containing static and time-varying components. The shaped pulse is employed to dynamically correct the errors caused by the noises above. (b) Detection of the $^{13}$C-nuclear-spin reservoir with the NV spin coherence under a DD-32 sequence. The horizontal axis is the interval $\tau$ between two $\pi$ pulses (see Supplemental Material~\cite{sm}). The signatures of the proximal five $^{13}$C spins (dips 1-5) emerge around the ninth resonance dip (the order $k = $ 9) corresponding to the interval $\tau = (2k-1)\pi/\omega_{\text{C}}$, where $\omega_{\text{C}}$ is the $^{13}$C Larmor precession frequency under a magnetic field of $\approx$ 510 G. The dip 1$^{\prime}$ belongs to $k = $ 8, while the dips 2$^{\prime}$ and 3$^{\prime}$ belong to $k = $ 10.  The solid line represents the coherence curve calculated based on the coupling parameters from (c). (c) The coupling parameters of the five $^{13}$C spins by fitting the results with different orders under the DD-32 sequence and the accordingly calculated noise spectrum under a magnetic field of about 510 G. $A_{\parallel}$ and $A_{\perp}$ represent the hyperfine components $A_{zz}$ and $\sqrt{A_{zx}^2+A_{zy}^2}$. All errors in parentheses are one standard deviation.
}\label{noise}
\end{figure}

As illustrated in Fig.~\ref{noise}(a), the NV center is formed by a substitutional nitrogen atom and an adjacent vacancy in diamond lattice. The diamond used in this work is prepared with $^{13}$C natural abundance (1.1\%) (see Supplemental Material~\cite{sm}\nocite{xie2021identity, khaneja2005optimal, pedersen2007fidelity}), which brings two advantages: one is that strongly coupled $^{13}$C spins are retained and available for extra quantum resources~\cite{waldherr2014quantum, bradley2019ten, xie2021beating}; the other is avoiding costly $^{12}$C isotope purification~\cite{balasubramanian2009ultralong}. The spin coherence of the NV electron is heavily disturbed by three kinds of noise sources existing in diamond: the electron spins owned by lattice defects, $^{13}$C nuclear spins, and lattice oscillations. The noise from lattice defects is ignorable with the ultrapure diamond after high-temperature annealing. The effect of lattice oscillations is limiting the longitudinal relaxation time of the NV spin whose error is not resistable by DEC due to its Markovian behavior but can be reduced orders of magnitude at cryogenic temperatures~\cite{abobeih2018one}. Therefore, the noise considered to be resisted is mainly originated from the $^{13}$C nuclear spins that are coupled to the NV electron spin with magnetic dipolar interactions.

Different from the previous works where the $^{13}$C spin noise is simply taken as static~\cite{rong2015experimental, waldherr2014quantum, dolde2014high}, a complete noise model is constructed here by precisely measuring the $^{13}$C spin environment. Generally, the $^{13}$C spins in the reservoir are categorized into two groups, that is, the strongly coupled $^{13}$C spins in the proximity and the other weakly coupled $^{13}$C spins, as shown in Fig.~\ref{noise}(a). Five strongly coupled $^{13}$C spins is detected by monitoring the NV coherence under dynamical decoupling (DD) sequences~\cite{taminiau2012detection, abobeih2018one}, as shown in Fig.~\ref{noise}(b). The coupling parameters of the five $^{13}$C spins are obtained by fitting each resonance dip (see more details in the Supplemental Material~\cite{sm}), with their results gathered in the table of Fig.~\ref{noise}(c). In the weak coupling limit, the parallel component $A_{\parallel}$ of the $^{13}$C hyperfine interaction transitions into classical static noise, while the transverse component $A_{\perp}$ under the $^{13}$C spin precession performs like time-varying noise. Therefore, the residual noise from the weakly coupled $^{13}$C spins is treated as classical, including static and time-dependent components. The static component obeys a zero-mean Gaussian distribution with the strength measured to be $\sigma \approx$ 20 kHz. The time-varying part $\eta(t)$ has two quadrature amplitudes $X$ and $Y$ with the strengths measured to be $\sigma_x = \sigma_y \approx $ 30 kHz, and is given by $\eta(t) = X \cos(\omega_{\text{C}} t) + Y \sin(\omega_{\text{C}} t)$ with $\omega_{\text{C}}$ denoting the $^{13}$C Larmor precession frequency (see Supplemental Material~\cite{sm}). The five $^{13}$C spins are also classicized with the time-varying components displayed on the right of Fig.~\ref{noise}(c), for analyzing the noise-filtering behavior of the designed pulses in Fig.~\ref{resistance}(a).

With the noise model above, the shaped pulse is designed to cancel the errors arising therefrom for achieving a high-fidelity CNOT gate. The CNOT gate is the nuclear-state-controlled electron NOT (C$_n$NOT$_e$) gate performed upon the hybrid system constituted by the NV electron spin and the $^{14}$N nuclear spin. Considering the complexity of the problem, the numerical method is adopted to optimize the shaped pulse for filtering the noise with higher orders. It is worth noting that it is unnecessary to take the nearby $^{13}$C spins as quantum objects in the numerical optimization, which dramatically reduces the required computation overhead and make the optimized shaped pulse applicable for the other NVs without the need for a priori knowledge of the $^{13}$C spin distribution (see Supplemental Material~\cite{sm}).

In order to estimate the performance of the optimized pulse, the ability to resist different time-varying noises is firstly investigated~\cite{kabytayev2014robustness, huang2017robust}. The noise-filtering method, just as used in DD~\cite{biercuk2009optimized, ma2015resolving}, is employed here for analyzing the gate fidelity by calculating the overlap integral between the filter function $F(\omega)$ and the noise spectrum $S(\omega)$. Nevertheless, due to the control-noise noncommutativity, the expression of the gate fidelity must be expanded in series~\cite{green2012high, paz2014general}
\begin{equation}
\begin{aligned}
\label{filtering}
\mathcal{F}_g = 1 &- \frac{1}{2\pi} \int_{-\infty}^{+\infty} \frac{d\omega}{\omega^2} S(\omega) F_1(\omega) \\
&- \frac{1}{(2\pi)^2} \int_{-\infty}^{+\infty} \frac{d\omega}{\omega^2} S(\omega) \int_{-\infty}^{+\infty} \frac{d\omega^\prime}{{\omega^\prime}^2} S(\omega^\prime) F_2(\omega,\omega^\prime) - \cdots
\end{aligned}
\end{equation}
where $F_1(\omega)$ and $F_2(\omega,\omega^\prime)$ are the filter functions of the first two orders. Since the noise-filtering behavior of the optimized pulse can not be sufficiently depicted by the leading-order term, we directly calculated the gate errors under different time-varying noises, which can serve as an all-order filter function in some sense. The results obtained for the optimized pulse with a duration of 1.5 \textmu s are plotted in Fig.~\ref{resistance}(a), together with those of the primitive pulse (401 ns) and another shaped pulse with only static-noise resistance (1.5 \textmu s) for comparison. The latter performs badly near the $^{13}$C Larmor frequency $\omega_{\text{C}}$, and even worse than the primitive pulse due to a longer duration, which justifies constructing the complete noise model instead of a simplified static one~\cite{rong2015experimental, waldherr2014quantum, dolde2014high}. The time-varying-noise resistance near $\omega_{\text{C}}$ optimized for the shaped pulse also works for the proximal five $^{13}$C spins with all errors near the minimum below $10^{-4}$, as shown in Fig.~\ref{resistance}(b). The performances of the pulses above manifest even more obvious by monitoring the state evolutions under 1000 noise samples at $\omega_{\text{C}}$, graphically represented as the trajectories of the Bloch vector in Fig.~\ref{resistance}(c). The trajectories for the time-varying-noise-resistant pulse finally converge into a small region, in comparison to the largely dispersed regions exhibited by the other two pulses. As a matter of fact, the effect of the strongly coupled $^{13}$C spins can be precisely analyzed by absorbing them to constitute a seven-qubit quantum register, with their locations given in Fig.~\ref{resistance}(d). Under this scenario, the quantum-state purity is utilized to evaluate the ability to resist the quantum noise, or say the crosstalk, from other qubits, as displayed in Fig.~\ref{resistance}(e). The final purity is recovered above 0.9999 for the shaped pulse with time-varying noise optimized, which signifies that destructive entanglements with the $^{13}$C spins are cancelled out among the subpulses of the shaped pulse.

\begin{figure}
\centering
\centering
\begin{overpic}[width=0.95\columnwidth]{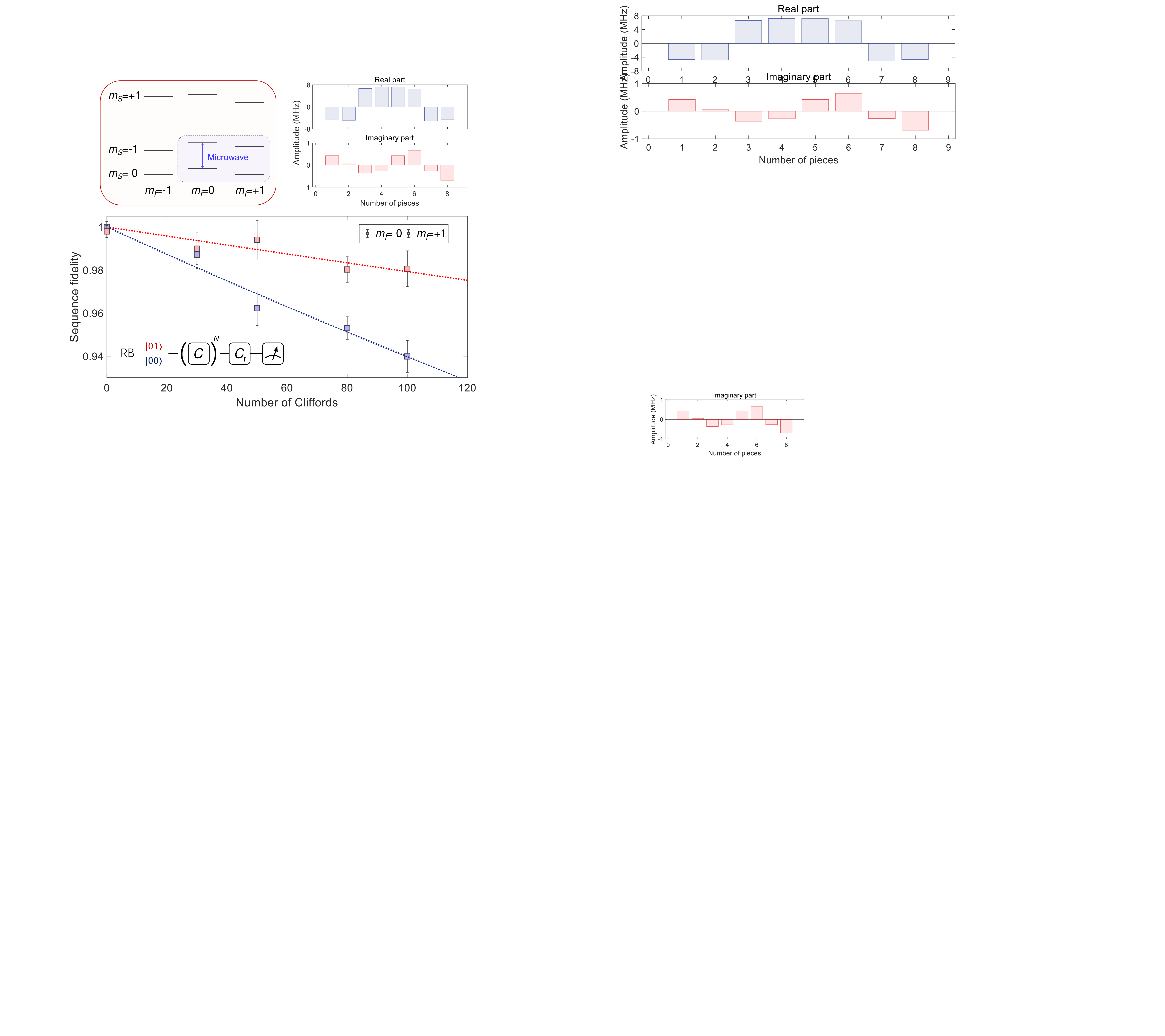}
	\put (0., 80.2) {\figfont{(a)}}
	\put (54.4, 80.2) {\figfont{(b)}}
	\put (0., 47.2) {\figfont{(c)}}
\end{overpic}
\caption{Single-qubit gates and randomized benchmarking. (a) Level diagram of the NV-$^{14}$N coupled system. The four levels inside the shaded box are utilized as two qubits and the microwave is resonantly applied in the spin state $m_{\scaleto{I}{4.5pt}} = 0$. (b) The optimized shaped pulse that simultaneously performs the $\pi/2$ gate in both $m_{\scaleto{I}{4.5pt}} = 0$ and $m_{\scaleto{I}{4.5pt}} = +1$ subspaces. The shaped pulse exhibits a piecewise-constant amplitude (Rabi frequency) profile with 8 pieces of 30-ns subpulses in both real and imaginary parts. (c) The single-qubit RB results for two nuclear spin states. The Clifford gates are comprised of the $\pi/2$ gate in (b). By fitting the results, the average fidelities per Clifford are given by 99.936(4)\% for the state $m_{\scaleto{I}{4.5pt}} = 0$ and 99.979(3)\% for the state $m_{\scaleto{I}{4.5pt}} = +1$. Inset: the implemented RB sequences.
}\label{single}
\end{figure}

\begin{figure}
\centering
\begin{overpic}[width=1.0\columnwidth]{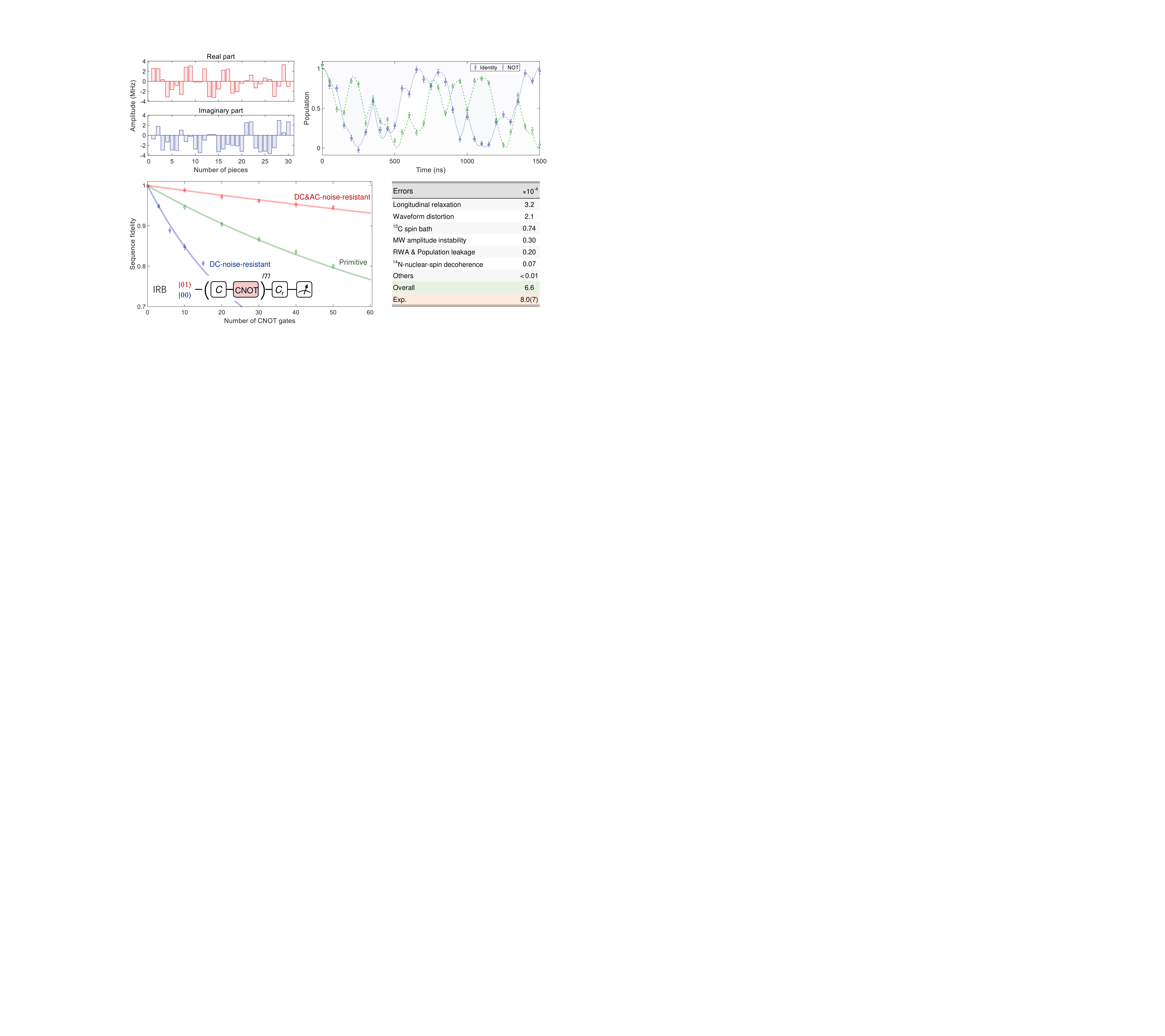}
	\put (0., 64.1) {\figfont{(a)}}
	\put (41.5, 64.1) {\figfont{(b)}}
	\put (0., 35.3) {\figfont{(c)}}
	\put (59.8, 35.3) {\figfont{(d)}}
\end{overpic}
\caption{CNOT gates and gate error analysis. (a) The optimized shaped pulse for realizing a high-fidelity CNOT gate. As shown in Fig.~\ref{resistance}, the shaped pulse is resistant to static noise, time-varying noise, and quantum noise. It has 30 pieces of 50-ns subpulses in both real and imaginary parts. (b) The measured evolutions under the NOT and Identity operations of the CNOT gate by applying the shaped pulse in (a). The solid lines are the theoretical simulations. (c) The fidelity measurement of the CNOT gates realized by the pulses in Fig.~\ref{resistance} by the modified IRB. Inset: the implemented modified IRB sequences. (d) Error budget for the residual gate infidelity.
 }\label{CNOT}
\end{figure}

In the following, the high fidelity manifested by the shaped pulse optimized above is experimentally characterized by the Clifford-based randomized benchmarking (RB)~\cite{magesan2011scalable, magesan2012efficient}. However, it is inconvenient to directly apply two-qubit RB for the hybrid system here, due to the control difference between the electron spin and the nuclear spin. Alternatively, considering the negligible error from nuclear-spin decoherence (Fig.~\ref{CNOT}(d)), the two-qubit RB for the C$_n$NOT$_e$ gate can be simplified into estimating the fidelities of the Identity operation and the NOT operation in two nuclear subspaces, which is called subspace randomized benchmarking~\cite{baldwin2020subspace} (see Supplemental Material~\cite{sm}). Before benchmarking the operations above by interleaved RB (IRB)~\cite{magesan2012efficient}, single-qubit RB needs to be implemented first to serve as a reference. The shaped pulse shown in Fig.~\ref{single}(b) is designed for performing the $\pi/2$ gate in both subspaces defined as Fig.~\ref{single}(a), from which single-qubit Cliffords are established. The average fidelities per Clifford are measured to be 99.936(4)\% for the state $m_{\scaleto{I}{4.5pt}} = 0$ and 99.979(3)\% for the state $m_{\scaleto{I}{4.5pt}} = +1$ by fitting the RB results in Fig.~\ref{single}(c). Since every Clifford has about 2.2 $\pi/2$ gates on average, the fidelity of the $\pi/2$ gate turns out to be 99.980(1)\%. As described in Fig.~\ref{resistance}, the shaped pulse with the profile (Fig.~\ref{CNOT}(a)) is optimized to resist the complete noise model, including static and time-varying, classical and quantum noises. The noise-resistant ability is acquired by experiencing the complicated evolutions displayed in Fig.~\ref{CNOT}(b), corresponding to the Identity operation and the NOT operation in both subspaces, respectively. By applying the IRB sequences, the fidelity of the CNOT gate realized by the shaped pulse is found to be 99.920(7)\% in Fig.~\ref{CNOT}(c) after subtracting the single-qubit reference. Compared with the primitive gate with a fidelity of 99.52(2)\%, the shaped pulse with only static-noise resistance has a lower fidelity of 98.27(6)\% for the fact that the noise model is inaccurate and a longer gate time leads to a larger error, as stated regarding DEC.

The residual error 0.80(7)\textperthousand\ is accountable by scrutinizing the errors from varieties of experimental imperfections that are collected in Fig.~\ref{CNOT}(d) (see a detailed analysis in the Supplemental Material~\cite{sm}). After suppressing the error from the spin reservoir by two orders of magnitude, the errors from the longitudinal relaxation of the NV spin and the waveform distortion of the shaped pulse dominate. The error from the longitudinal relaxation is estimated to be 0.32\textperthousand\ by measuring three-level relaxation behaviors, which can be removed by running at cryogenic temperatures~\cite{abobeih2018one}. The error from the waveform distortion is 0.21\textperthousand\ by performing the state tomography after the IRB sequences, and can be further mitigated by carefully refining the waveform details or by quantum optimal control~\cite{li2017hybrid, wierichs2022general}. The remanent error from the $^{13}$C spin bath is calculated to be 0.074\textperthousand\ based on the noise model above, including the five$^{13}$C spins, the static noise (20 kHz), and the time-varying noise (30 kHz). Other errors are insignificant: the amplitude instability of the applied microwave produces an error of 0.030\textperthousand; the effects from the rotation wave approximation (RWA) and the population leakage into the third level are calculated together to give an error of 0.020\textperthousand; the $^{14}$N-nuclear-spin decoherence from the applied pulse not from the $^{13}$C spin bath gives an error of $7\times10^{-6}$; the errors from other sources are all below $10^{-6}$.

In conclusion, we demonstrate a 99.920(7)\%-fidelity CNOT gate in a solid-state spin system by employing DEC. The spin reservoir is precisely measured for establishing a complete and real noise model, consisting of static noise, time-varying noise, and quantum noise. The shaped pulse is exquisitely designed to dynamically correct the errors from all kinds of noises. The method above is extensively applicable for solid-state spin systems , such as phosphorus dopants~\cite{morton2008solid, mkadzik2022precision} and quantum dots~\cite{noiri2022fast, xue2022quantum} in silicon, defects in diamond and silicon carbide~\cite{widmann2015coherent, wolfowicz2021quantum}, rare-earth-doped crystals~\cite{siyushev2014coherent, ruskuc2022nuclear}, and so forth. The analysis of the residual gate error indicates that all errors above $10^{-4}$ can be removed technically, providing the possibility to improve the fidelity up to 99.99\%. The high-fidelity CNOT gate realized here is of significant importance for practical quantum computation. Furthermore, considering being performed on a hybrid system, it is also crucial for exchanging quantum information on the interface of quantum memory~\cite{morton2008solid, maurer2012room}.

\section*{Acknowledgements}
We thank Jianpei Geng (University of Stuttgart, Germany) for helpful discussions. This work was supported by the National Key R\&D Program of China (Grant No. 2018YFA0306600, 2016YFA0502400), the National Natural Science Foundation of China (Grant No. 81788101, 91636217, T2125011, 12274396), Innovation Program for Quantum Science and Technology (Grant No. 2021ZD0302200, 2021ZD0303204), the CAS (Grant No. XDC07000000, GJJSTD20200001, QYZDY-SSW-SLH004, Y201984), the Anhui Initiative in Quantum Information Technologies (Grant No. AHY050000), the Fundamental Research Funds for the Central Universities, the China Postdoctoral Science Foundation (Grant No. 2021M703110, 2022T150631), and the Hefei Comprehensive National Science Center.

\clearpage

\bibliography{references.bib}

\clearpage

\clearpage

\begin{figurehere}
\centering
\begin{overpic}[width=1.0\columnwidth]{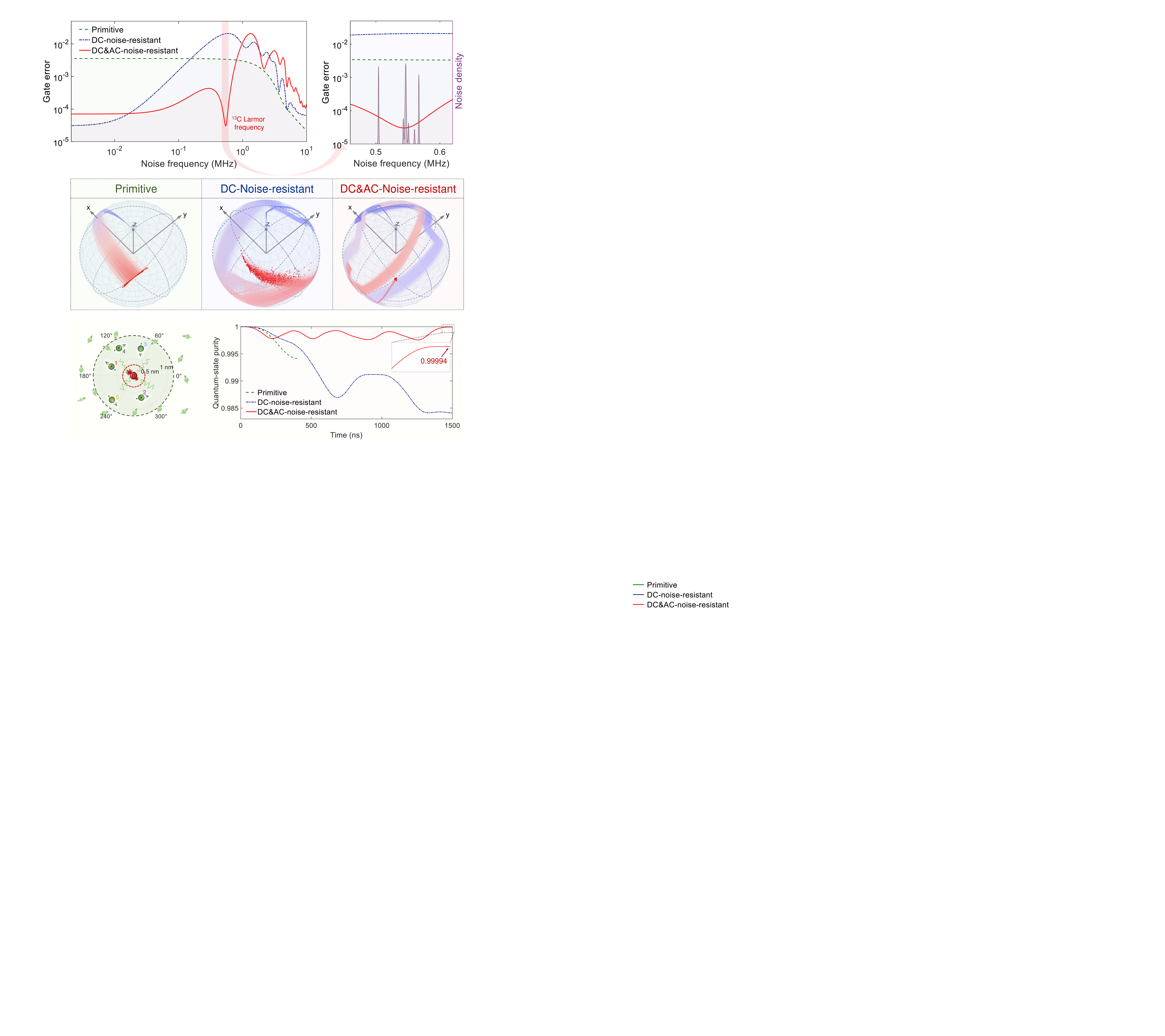}
	\put (0.0, 99.8) {\figfont{(a)}}
	\put (65.8, 99.8) {\figfont{(b)}}
	\put (3, 62.2) {\figfont{(c)}}
	\put (4, 27.2) {\figfont{(d)}}
	\put (39.8, 27.2) {\figfont{(e)}}
\end{overpic}
\caption{Resistance to time-dependent noise and quantum noise. (a) The CNOT gate error induced by time-dependent noise for three kinds of control pulses. The primitive gate is realized by a rectangular pulse (green), while the other two are constructed by shaped pulses to dynamically correct the noise-induced errors, of which the one is optimized to resist only static noise (blue) and the other is resistant to both static and time-varying noises (red). In the calculation, two components of the time-dependent noise obey Gaussian distributions with the strengths $\sigma_x = \sigma_y = $ 70 kHz. (b) The zoom-in display near the $^{13}$C Larmor frequency in (a), together with the $^{13}$C noise spectrum (Fig.~\ref{noise}(c)). (c) The state evolution under the NOT operation of the CNOT gate with 1000 samples of the time-varying noise at the $^{13}$C Larmor frequency. (d) Localization of the proximal five $^{13}$C spins that function as quantum noise in (e). (e) Resistance to quantum noise exhibited by the quantum-state purity under the NOT operation of the CNOT gate. The five $^{13}$C spins are taken as quantum objects and and their couplings to the NV center are included in the whole Hamiltonian. See the behaviors under the Identity operation of the CNOT gate in the Supplemental Material~\cite{sm}.
}\label{resistance}
\end{figurehere}

\clearpage

\clearpage

\end{document}

\title{Supplemental Material for \\ 99.92\%-Fidelity CNOT Gates in Solids by Filtering Time-dependent and Quantum Noises}

\author{Tianyu Xie}
\thanks{T. Xie and Z. Zhao contributed equally to this work.}
\affiliation{CAS Key Laboratory of Microscale Magnetic Resonance and School of Physical Sciences, University of Science and Technology of China, Hefei 230026, China}
\affiliation{CAS Center for Excellence in Quantum Information and Quantum Physics, University of Science and Technology of China, Hefei 230026, China}

\author{Zhiyuan Zhao}
\thanks{T. Xie and Z. Zhao contributed equally to this work.}
\affiliation{CAS Key Laboratory of Microscale Magnetic Resonance and School of Physical Sciences, University of Science and Technology of China, Hefei 230026, China}
\affiliation{CAS Center for Excellence in Quantum Information and Quantum Physics, University of Science and Technology of China, Hefei 230026, China}

\author{Shaoyi Xu}
\affiliation{CAS Key Laboratory of Microscale Magnetic Resonance and School of Physical Sciences, University of Science and Technology of China, Hefei 230026, China}
\affiliation{CAS Center for Excellence in Quantum Information and Quantum Physics, University of Science and Technology of China, Hefei 230026, China}

\author{Xi Kong}
\affiliation{National Laboratory of Solid State Microstructures and Department of Physics, Nanjing University, Nanjing 210093, China}

\author{Zhiping Yang}
\affiliation{CAS Key Laboratory of Microscale Magnetic Resonance and School of Physical Sciences, University of Science and Technology of China, Hefei 230026, China}
\affiliation{CAS Center for Excellence in Quantum Information and Quantum Physics, University of Science and Technology of China, Hefei 230026, China}

\author{Mengqi Wang}
\affiliation{CAS Key Laboratory of Microscale Magnetic Resonance and School of Physical Sciences, University of Science and Technology of China, Hefei 230026, China}
\affiliation{CAS Center for Excellence in Quantum Information and Quantum Physics, University of Science and Technology of China, Hefei 230026, China}

\author{Ya Wang}
\affiliation{CAS Key Laboratory of Microscale Magnetic Resonance and School of Physical Sciences, University of Science and Technology of China, Hefei 230026, China}
\affiliation{CAS Center for Excellence in Quantum Information and Quantum Physics, University of Science and Technology of China, Hefei 230026, China}
\affiliation{Hefei National Laboratory, University of Science and Technology of China, Hefei 230088, China}

\author{Fazhan Shi}
\email{fzshi@ustc.edu.cn}
\affiliation{CAS Key Laboratory of Microscale Magnetic Resonance and School of Physical Sciences, University of Science and Technology of China, Hefei 230026, China}
\affiliation{CAS Center for Excellence in Quantum Information and Quantum Physics, University of Science and Technology of China, Hefei 230026, China}
\affiliation{Hefei National Laboratory, University of Science and Technology of China, Hefei 230088, China}
\affiliation{School of Biomedical Engineering and Suzhou Institute for Advanced Research, University of Science and Technology of China, Suzhou 215123, China}

\author{Jiangfeng Du}
\email{djf@ustc.edu.cn}
\affiliation{CAS Key Laboratory of Microscale Magnetic Resonance and School of Physical Sciences, University of Science and Technology of China, Hefei 230026, China}
\affiliation{CAS Center for Excellence in Quantum Information and Quantum Physics, University of Science and Technology of China, Hefei 230026, China}
\affiliation{Hefei National Laboratory, University of Science and Technology of China, Hefei 230088, China}

\maketitle

\section*{Diamond sample}
The targeted NV center resides in a bulk diamond whose top face is perpendicular to the [100] crystal axis and lateral faces are perpendicular to the [110] axis. The bulk diamond is ultrapure with the nitrogen concentration less than 5 p.p.b., and the abundance of $^{13}$C atoms is at the natural level of 1.1$\%$. The diamond is irradiated using 10-MeV electrons to a total dose of $\sim$ 1.0 kgy. A solid immersion lens (SIL) is created around the targeted NV center to increase the luminescence rate of the NV to $\sim$ 300 kcounts/s.

\section*{Experimental setup}
\begin{figure}
\centering
\includegraphics[width=1.0\textwidth]{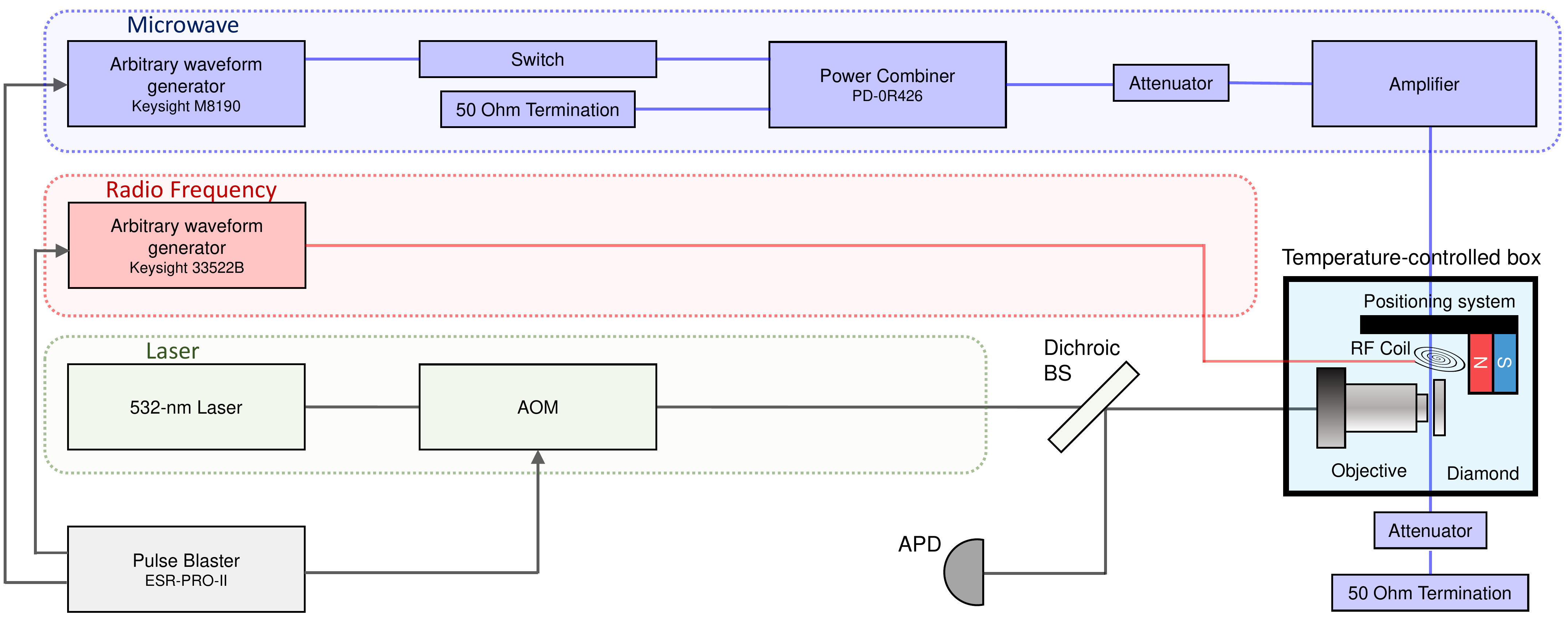}
\caption{Simplified experimental setup. The setup mainly consists of microwave (blue) and radio-frequency (red) circuits, optical systems (green), synchronization system, and a diamond platform inside a temperature-controlled box. 
}\label{setup}
\end{figure}

The diamond is mounted on a typical confocal setup to perform optically detected magnetic resonance, synchronized with a microwave bridge by a multichannel pulse blaster (Spincore, PBESR-PRO-500). The 532-nm green laser for driving NV electron dynamics, and sideband fluorescence (650–800 nm) go through the same oil objective (Olympus, UPLSAPO 100XO, NA 1.40). To protect the NV center's negative state and longitudinal relaxation time against laser leakage effects, all laser beams pass twice through acousto-optic modulators (AOM) (Gooch $\&$ Housego, power leakage ratio $\sim$ 1/1,000) before they enter the objective. The fluorescence photons are collected by avalanche photodiodes (APD) (Perkin Elmer, SPCM-AQRH-14) with a counter card (National Instruments, 6612). The 1.4 GHz and 4.3 GHz microwave (MW) pulses for manipulating three sublevels of the NV center are generated from an arbitrary waveform generator (AWG) (Keysight M8190), and fed into the coplanar waveguide microstructure. The 5.1 MHz radio-frequency (RF) pulses are generated from another AWG (Keysight 33522B), and applied for controlling the $^{14}$N nuclear spin through a RF coil. The external magnetic field ($\approx$ 510 G) is generated from a permanent magnet and aligned parallel to the NV axis through a three-dimensional positioning system. The positioning system, together with the platform holding the diamond and the objective, is placed inside a thermal insulation copper box. The temperature inside the copper box is stabilized at $\sim$ mK level through the feedback of the temperature controller (Stanford, PTC10). The overall setup is plotted in Fig.~\ref{setup}.

\clearpage

\section*{System Hamiltonian and noise Hamiltonian}
The whole Hamiltonian for realizing the high-fidelity CNOT gate in this work, contains the terms concerning the system and the environmental noise.
\begin{equation}
\label{hamiltonian}
H = H_{\text{sys}} + H_{\text{noi}}
\end{equation}
The system upon which the CNOT gate is performed, consists of the NV electron spin ($S=1$) and the $^{14}$N nuclear spin ($I=1$). A convention is made here that all physical quantities with the dimension of energy in the equations below are represented in the dimension of circular frequency, and their values are given in the dimension of frequency.

With an external magnetic field $B_0 \approx$ 510 G applied along the axis of the NV center (defined as the $z$ axis), the system Hamiltonian is given by
\begin{equation}
\label{system}
H_{\text{sys}} = D S_z^2 + \gamma_e B_0 S_z + Q (I_z^{\text{N}})^2 + \gamma_{\text{N}} B_0 I_z^{\text{N}} + A_{\parallel} S_z I_z^{\text{N}} + A_{\perp} (S_x I_x^{\text{N}}  + S_y I_y^{\text{N}})
\end{equation}
where $D = $ 2869.901 MHz are the zero-field splitting of the NV electron spin. $\gamma_e$ and $\gamma_{\text{N}}$ are the gyromagnetic ratios of the NV spin and the $^{14}$N spin, while $S_{x/y/z}$ and $I_{x/y/z}^{\text{N}}$ denote the $x/y/z$ components of the spin operators for two spins. By means of the method in the work~\cite{xie2021identity}, the $^{14}$N quadrupole coupling $Q$, the parallel component $A_{\parallel}$, the transverse component $A_{\perp}$ of the hyperfine interaction, and the ratio $\gamma_e/\gamma_{\text{N}}$ are given by -4945.8003(12) kHz, -2164.6714(17) kHz, -2632.8(8) kHz, and -9113.77(8), respectively. As described in the text, the system is disturbed by interactions with $N$ $^{13}$C nuclear spins
\begin{equation}
\label{noise_13C}
H_{\text{noi}}^{N} = \sum_{i=1}^{N} \gamma_{\text{C}} B_0 I_z^{i} + S_z (A_{zx}^i I_x^{i}  + A_{zy}^i  I_y^{i} + A_{zz}^i I_z^{i})
\end{equation}
where the $^{13}$C Larmor frequency $\omega_{\text{C}} = \gamma_{\text{C}} B_0 \approx$ -546.67 kHz. Since the hyperfine interactions are weak enough, Eq.~(\ref{noise_13C}) excludes non-secular coupling terms, which mix the states of the NV electron spin and thus are negligible here. In the interaction picture of the $^{13}$C spin precession, the noise Hamiltonian becomes time-dependent
\begin{equation}
\label{noise_13C_time-varying}
\begin{aligned}
H_{\text{noi}}^{N}(t) = & \left( \cos(\omega_{\text{C}}t) \sum_{i=1}^{N} (A_{zx}^i I_x^{i} + A_{zy}^i I_y^{i}) + \sin(\omega_{\text{C}}t) \sum_{i=1}^{N} (A_{zx}^i I_y^{i} - A_{zy}^i I_x^{i}) \right) S_z\\
&+ S_z \sum_{i=1}^{N} A_{zz}^i I_z^{i}
\end{aligned}
\end{equation}
In the weak coupling limit, the overall effect of a great number of $^{13}$C spins is like classical noise, and that is, three sums in Eq.~(\ref{noise_13C_time-varying}) become three Gaussian distributions based on the central limit theorem. Therefore, apart from the proximal five $^{13}$C spins detected by DD sequences as shown in Fig.~1(b) of the main text, the other $^{13}$C spins can be taken as a whole classical noise. By replacing the sums of Eq.~(\ref{noise_13C_time-varying}) with three Gaussian random variables $X$, $Y$, and $Z$, the $^{13}$C spin bath is modelled by
\begin{equation}
\label{noise_model}
H_{\text{noi}}(t) = (X \cos(\omega_{\text{C}} t) + Y \sin(\omega_{\text{C}} t) + Z) S_z + H_{\text{noi}}^{5}
\end{equation}
where $H_{\text{noi}}^{5}$ is the Hamiltonian of the five $^{13}$C spins in the Schr\"{o}dinger picture, given by Eq.~(\ref{noise_13C}). The time-dependent part $X \cos(\omega_{\text{C}} t) + Y \sin(\omega_{\text{C}} t)$ in Eq.~(\ref{noise_model}) is exactly the time-varying noise $\eta(t)$ used in the main text, while $Z$ represents the static noise. In this work, the gate error induced by the $^{13}$C spin bath is estimated based on the model above, but the numerical optimization of the shaped pulse for acquiring the high-fidelity CNOT gate takes the five$^{13}$C spins as classical as well, as stated in the main text. It can be inferred from the deduction above that the noise model used here is close enough to the real noise environment.

After abandoning the spin state $|{m_{\scaleto{S}{4.5pt}} = +1 \rangle}$ of the NV electron spin and the state $|{m_{\scaleto{I}{4.5pt}} = -1 \rangle}$ of the $^{14}$N nuclear spin, the left four levels comprise the two qubits used for realizing the CNOT gate, as shown in Fig.~3(a) of the main text. The system Hamiltonian in Eq.~(\ref{system}) reduced to the four-level subspace after removing some constant terms is given by
\begin{equation}
\label{system_reduced}
H_{\text{sys}} = \omega_{\scaleto{S}{4.5pt}} S_z + \omega_{\text{N}} I_z^{\text{N}} + A_{\parallel} S_z I_z^{\text{N}}
\end{equation}
where $S_z$ and $I_z^{\text{N}}$ are changed into the spin-1/2 operators for two spins, and $\omega_{\scaleto{S}{4.5pt}}$ and $\omega_{\text{N}}$ are the transition frequencies of two spins. The coupling term $A_{\parallel}$ is adjusted to be 2159.876 kHz for the four levels above resulting from the second perturbative effect of the transverse component $A_{\perp}$. Accordingly, the form of the hyperfine interactions with the surrounding $^{13}$C spins needs to be changed, especially for $H_{\text{noi}}^{5}$ in the noise model above
\begin{equation}
\label{noise_reduced}
\begin{aligned}
H_{\text{noi}}^{5} &= \sum_{i=1}^{5} \omega_{\text{C}} I_z^{i} + (S_z - \frac{\mathbb{I}}{2}) (A_{zx}^i I_x^{i}  + A_{zy}^i  I_y^{i} + A_{zz}^i I_z^{i})\\
&\approx \sum_{i=1}^{5} \omega_{\text{C}}^i I_z^{i} + S_z (A_{zx}^i I_x^{i}  + A_{zy}^i  I_y^{i} + A_{zz}^i I_z^{i})
\end{aligned}
\end{equation}
where $\omega_{\text{C}}^i = \omega_{\text{C}} - A_{zz}^i/2$. The second line in Eq.~(\ref{noise_reduced}) abandons the terms $-A_{zx}^i I_x^i/2$ and $-A_{zy}^i I_y^i/2$, since $A_{zx}^i$ and $A_{zy}^i$ are much smaller than $\omega_{\text{C}}$. For the strongly coupled $^{13}$C spins, their precession frequencies are slightly changed by the hyperfine interactions with the NV center, which explains the shifted noise peaks of the five $^{13}$C spins in Fig.~1(c) and Fig.~2(b) of the main text.

As described in Fig.~3(a), the MW pulse is resonantly applied in the nuclear state $m_{\scaleto{I}{4.5pt}} = 0$ with the frequency $\omega_{\text{MW}} = \omega_{\scaleto{S}{4.5pt}} - A_{\parallel}/2$. Taking the interaction picture with respect to $\omega_{\text{MW}} S_z$ and $(\omega_{\text{N}} - \omega_{\text{sh}}) I_z$, the system Hamiltonian reads
\begin{equation}
\label{system_interaction}
H_{\text{sys}}^{\text{int}} = A_{\parallel} S_z (I_z^{\text{N}} + \frac{\mathbb{I}}{2}) + \omega_{\text{sh}} (I_z^{\text{N}} - \frac{\mathbb{I}}{2})
\end{equation}
where $\omega_{\text{sh}} = 1/(4T_{\text{CNOT}})$ with $T_{\text{CNOT}}$ denoting the duration of the realized CNOT gate. The constant term $- \omega_{\text{sh}} \mathbb{I}/2$ is added to compensate for the negative determinant of the CNOT gate. In the same picture, according to rotating wave approximtaion (RWA), the non-resonant component of the applied MW is abandoned. The Hamiltonian obtained for MW control is written as
\begin{equation}
\label{MW_control}
H_{\text{c}}(t) = \Omega_{\text{r}}(t) S_x + \Omega_{\text{i}}(t) S_y
\end{equation}
with arbitrarily adjustable $\Omega_{\text{r}}(t)$ and $\Omega_{\text{i}}(t)$ by time-dependently setting the amplitude and phase of the MW pulse generated by the AWG. The control Hamiltonian of Eq.~(\ref{MW_control}) together with the system Hamiltonian of Eq.~(\ref{system_interaction}) is used for generating the strictly defined CNOT gate under the noise model of Eq.~(\ref{noise_model}) by using $H_{\text{noi}}^{5}$ in Eq.~(\ref{noise_reduced}). The approximations in the derivation above, such as the reduction of the system Hamiltonian (population leakage) and RWA, are considered in the gate error analysis, while the approximations concerning the interactions with the $^{13}$C reservoir, at most, cause a small deviation in estimating the error from the $^{13}$C spin bath.

\section*{Experimental construction of the noise model}
\begin{figure}
\centering
\begin{overpic}[width=1.0\textwidth]{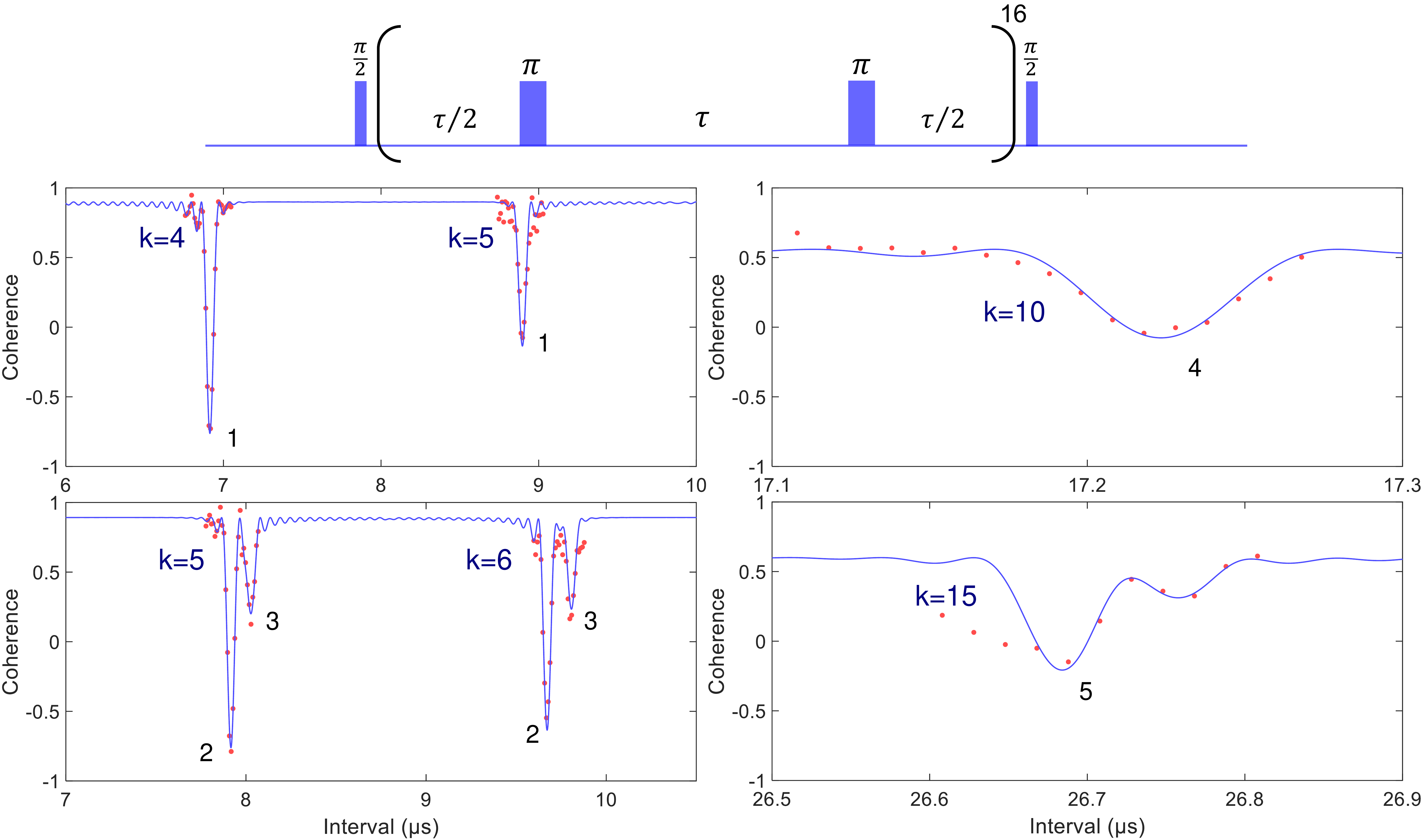}
\renewcommand{\familydefault}{\sfdefault}
	\put (13., 57.) {\figfont{(a)}}
	\put (0., 46.) {\figfont{(b)}}
	\put (50., 46.) {\figfont{(c)}}
	\put (0., 24.) {\figfont{(d)}}
	\put (50., 24.) {\figfont{(e)}}
\end{overpic}
\caption{Detection of the strongly coupled $^{13}$C spins with the NV spin coherence under DD sequences. (a) The DD sequence with 32 $\pi$ pulses (DD-32). The interval between two $\pi$ pulses is denoted by $\tau$. (b-e) Measured coherence under the DD-32 sequence with different orders $k$ for acquiring the coupling parameters of the five $^{13}$C spins in the table of Fig.~1(c). The horizontal axis is the interval $\tau$ between two $\pi$ pulses in (a). The coupling parameters of the first $^{13}$C spin are obtained by fitting two dips ($k = $ 4 and 5) in (b) with Eq.~(\ref{coherence_DD}). The solid line is the fitting curve plotted for comparison. The same procedures are performed for the other four $^{13}$C spins with the results of (c-e).
}\label{C13}
\end{figure}

All parameters in the noise model described by Eq.~(\ref{noise_model}) are determined experimentally. First of all, the coupling parameters of the nearby five $^{13}$C spins are obtained by adopting the method in the work~\cite{taminiau2012detection}. The DD sequences in Fig.~1(b) and Fig.~\ref{C13} are applied between the spin state $|{m_{\scaleto{S}{4.5pt}} = 0 \rangle}$ and the state ${| m_{\scaleto{S}{4.5pt}} = +1 \rangle}$ with the $\pi$ pulse interval $\tau$. It is assumed that a $^{13}$C spin coupled to the NV center has the coupling parameters $(A_{zx}, A_{zy}, A_{zz})$ as in Eq.~(\ref{noise_13C}). If the NV spin stays in the state $|{ m_{\scaleto{S}{4.5pt}} = 0 \rangle}$, the $^{13}$C spin feels the same field as the external field with the transition frequency $\omega_0=\omega_{\text{C}}$ and the direction denoted by $\hat{\mathbf{n}}_0$; if the NV spin stays in the state $|{m_{\scaleto{S}{4.5pt}} = +1 \rangle}$, the $^{13}$C spin is influenced by both the external field and the hyperfine interation, giving the resulting frequency $\omega_1=\sqrt{A_{zx}^2+A_{zy}^2+(\omega_{\text{C}}+A_{zz})^2}$ and the direction $\hat{\mathbf{n}}_1$. After some algebraic calculation, the decoherence of the NV spin may be caused by the non-parallelism between $\hat{\mathbf{n}}_0$ and $\hat{\mathbf{n}}_1$ with $| \hat{\mathbf{n}}_0 \cdot \hat{\mathbf{n}}_1 | < 1$
\begin{equation}
\label{coherence_DD}
C(\tau) = 1 - 2 \left(1 - (\hat{\mathbf{n}}_0 \cdot \hat{\mathbf{n}}_1)^2 \right) \sin^2(\frac{\omega_0 \tau}{4}) \sin^2(\frac{\omega_1 \tau}{4}) \frac{\sin^2(\frac{N \phi(\tau)}{2})}{\cos^2(\frac{\phi(\tau)}{2})}
\end{equation}
with
\begin{equation}
\label{coherence_phi}
\cos(\phi(\tau)) = \cos(\frac{\omega_0 \tau}{2}) \cos(\frac{\omega_1 \tau}{2}) - \hat{\mathbf{n}}_0 \cdot \hat{\mathbf{n}}_1 \sin(\frac{\omega_0 \tau}{2}) \sin(\frac{\omega_1 \tau}{2})
\end{equation}
The decoherence dip caused by the $^{13}$C spin will emerge repeatedly when the interval $\tau$ is equal to
\begin{equation}
\label{coherence_tau}
\tau_k = \frac{(2k-1)\pi}{(\omega_0+\omega_1)/2}
\end{equation}
with the resonance order $k$. According to the behavior of DD filter function~\cite{biercuk2009optimized, ma2015resolving}, a larger $\tau$ leads to a better spectral resolution, and thus a larger $k$ is preferred. The order $k = $ 9 is chosen to clearly distinguish the five $^{13}$C spins, as displayed in Fig.~1(b) of the main text. The coupling parameters are extracted by data fitting using Eq.~(\ref{coherence_DD}) with suitable orders $k$ for each $^{13}$C spin as shown in Fig.~\ref{C13}(b-e). The fitting results give the values of $A_{zz}$ and $\sqrt{A_{zx}^2 + A_{zy}^2}$, all gathered in the table of Fig.~1(c). It is not necessary to obtain the values of $A_{zx}$ and $A_{zy}$ for the error analysis of the CNOT gate, and thus an angle $\phi$ is arbitrarily chosen to assign these two values.

\begin{figure}
\centering
\begin{overpic}[width=1.0\textwidth]{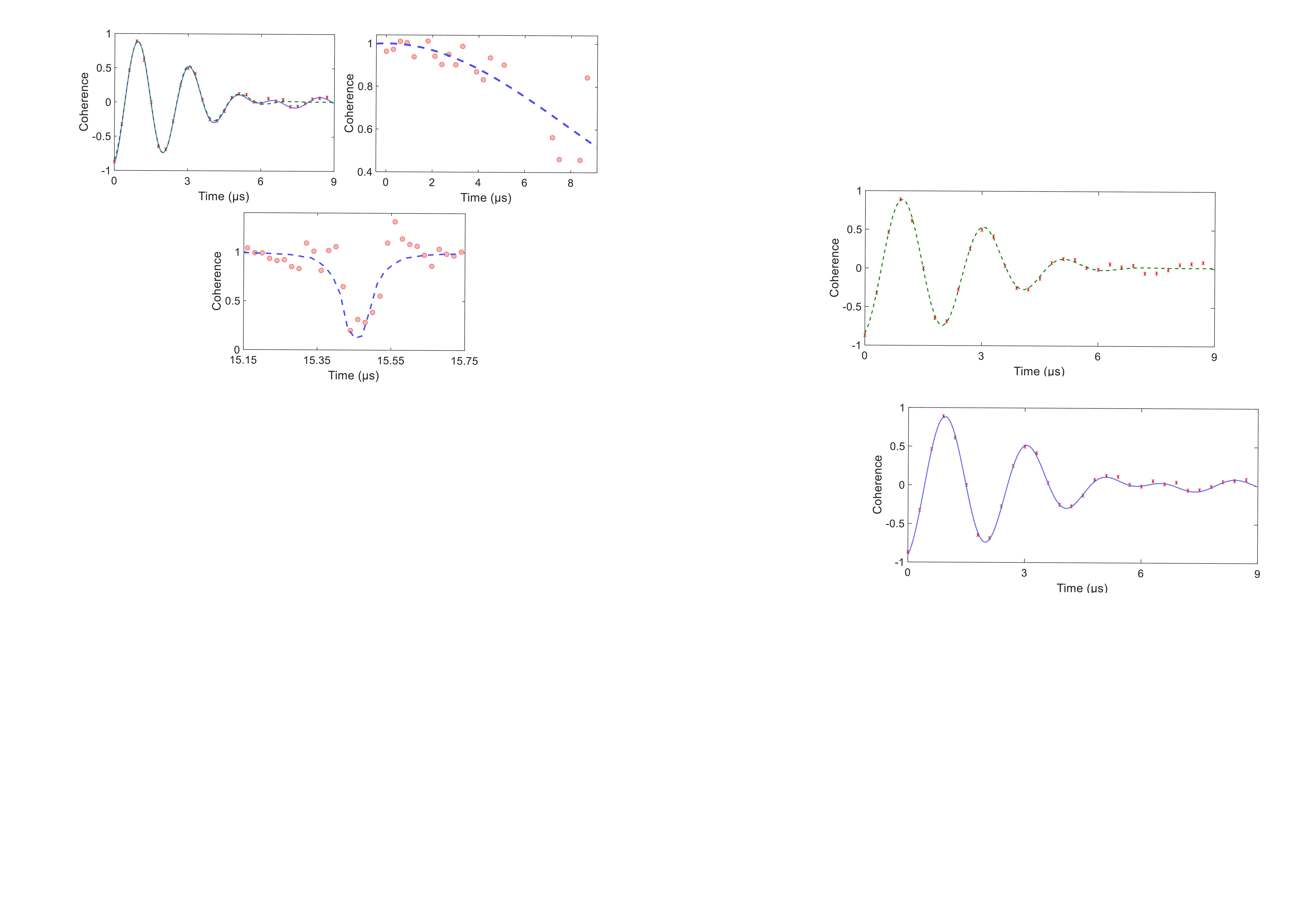}
\renewcommand{\familydefault}{\sfdefault}
	\put (0.5, 67.3) {\figfont{(a)}}
	\put (51., 67.3) {\figfont{(b)}}
	\put (25.5, 32.5) {\figfont{(c)}}
\end{overpic}
\caption{Measurement of the weakly coupled $^{13}$C spins taken as classic noise. (a) The NV spin coherence measured under a Ramesy sequence. The green line is data fitting using the function $a\cos(2\pi(\delta f)t+\phi_0)e^{-(t/T_2^*)^p}+b$ with $\delta f$ denoting the detuning. The blue line is plotted based on a numerical simulation, where the nearby five $^{13}$C spins and the static noise from the weakly coupled $^{13}$C spins (obtained in (b)) are considered. (b) The resultant decoherence from the weakly coupled $^{13}$C spins by deducting the contribution of the nearby five $^{13}$C spins from (a). The strength of the static noise from the weakly coupled $^{13}$C spins is obtained to be $\sigma = $ 20.0(1.5) kHz by fitting the results with the function $e^{-(2\pi\sigma t)^2/2}$. (c) The resonance dip of the weakly coupled $^{13}$C spins under the 32-DD sequence, obtained from the dip indicated by the label ``$^{13}$C Bath'' in Fig.~1(b) of the main text after subtracting the contribution of the nearby five $^{13}$C spins. The dashed line is the numerical simulation under the time-varying noise with the strengths $\sigma_x = \sigma_y = $ 30 kHz.
}\label{T2star}
\end{figure}

The other $^{13}$C spins are indistinguishable by DD sequences at room temperature, and thus treated as classical noise. Actually, the nearby five $^{13}$C spins can be taken as classical noise as well, because as described in the main text, they are not included in the numerical optimization but the shaped pulse optimized still bears the ability to disentangle with these $^{13}$C spins. From this aspect, it is justified that the other $^{13}$C spins are taken as a whole classical noise containing static and time-varying components. In the following, the strengths of the static noise and the time-varying noise, i.e., the standard deviations of three zero-mean Gaussian random variables $X$, $Y$, and $Z$ in Eq.~(\ref{noise_model}), are both determined. As shown in Fig.~\ref{T2star}(a), the NV coherence is measured under a Ramsey sequence. An exponential fit with the function $a\cos(2\pi(\delta f)t+\phi_0)e^{-(t/T_2^*)^p}+b$ can not explain the last few points very well, which manifests the signatures of the nearby five $^{13}$C spins. The contribution of the five $^{13}$C spins under the Ramsey sequence is numerically calculated and deducted from Fig.~\ref{T2star}(a), giving the results in Fig.~\ref{T2star}(b). Thus, the decoherence from the weakly coupled $^{13}$C spins is obtained. Since the effect of the time-varying noise is negligible for the NV coherence under the Ramsey sequence, the strength of the static noise is extracted to be $\sigma = $ 20.0(1.5) kHz by fitting with the function $e^{-(2\pi\sigma t)^2/2}$. As for the time-varying noise, the strengths of two quadrature amplitudes $X$ and $Y$ are determined by comprehensively considering the simulation of the dip indicated by the label ``$^{13}$C Bath'' in Fig.~1(b) of the main text and the gate fidelities calculated for the primitive gate and the shaped pulse with only static-noise resistance. It is known from Eq.~(\ref{noise_13C_time-varying}) that both random variables must have the same standard deviation, i.e., $\sigma_x = \sigma_y$. When the strengths $\sigma_x$ and $\sigma_y$ take the value of 30 kHz, the calculated fidelities for these two gates are 99.46\% and 98.26\%, agreeing well with the experimentally measured values 99.52(2)\% and 98.27(6)\% in the main text, and at the same time, the dip described above can be explained well by the simulation as shown in Fig.~\ref{T2star}(c).

In summary, all parameters in the noise model used in this work are determined by performing multiple experiments, consisting of the proximal five $^{13}$C spins with the coupling parameters listed in the table of Fig.~1(c), the static noise with the strength of 20 kHz and the time-varying noise with the strength of 30 kHz. The noise model is employed to estimate the gate error from the $^{13}$C spin bath and optimize the shaped pulse after transforming the five $^{13}$C spins into the corresponding static and time-varying noises.

\clearpage

\section*{Optimization of shaped pulses}
Considering the high computational overhead demanded for simulating five quantum objects ($^{13}$C spins), the noise model used for the optimization of shaped pulses has to be simplified at first. The five $^{13}$C spins are classicized in the manner of taking $A_{zz}/2$ as the strength of a static noise, and $\sqrt{A_{zx}^2 + A_{zy}^2}/2$ as the combined strength $\sqrt{\sigma_x^2+ \sigma_y^2}$ of a time-varying noise. The combined strength for each $^{13}$C spin is exactly the area of the corresponding peak in the noise spectrum of the five $^{13}$C spins as displayed in Fig.~1(c) of the main text. With the processing method above, the noise from the $^{13}$C spin bath is totally classical and has the form
\begin{equation}
\label{noise_optimization}
H_{\text{noi}}^{\text{opt}}(t) = (X^{\text{opt}} \cos(\omega_{\text{C}} t) + Y^{\text{opt}} \sin(\omega_{\text{C}} t) + Z^{\text{opt}}) S_z
\end{equation}
which is totally determined by new random variables $X^{\text{opt}}$, $Y^{\text{opt}}$, and $Z^{\text{opt}}$. Combined with the noise contributions from the weakly coupled $^{13}$C spins, the strengths of these random variables are given by 45.4 kHz, 45.4 kHz, and 53.4 kHz, respectively.

Combined with the system Hamiltonian of Eq.~(\ref{system_interaction}) and the control Hamiltonian of Eq.~(\ref{MW_control}), the unitary operation realized by a shaped pulse under a noise sample $x_i^{\text{opt}}$, $y_i^{\text{opt}}$, and $z_i^{\text{opt}}$, can be written as
\begin{equation}
\label{CNOT_noise}
U_i = \mathcal{T} e^{- i \int_0^{T_{\text{CNOT}}}(H_{\text{sys}}^{\text{int}} + H_{\text{c}}(t) + H_{i}^{\text{opt}}(t))dt}
\end{equation}
where $\mathcal{T}$ is the time ordering operator in that the integrand Hamiltonian is time-dependent. The noise Hamiltonian $H_{i}^{\text{opt}}(t)$ under the noise sample takes the form of $H_{i}^{\text{opt}}(t) = (x_i^{\text{opt}} \cos(\omega_{\text{C}} t) + y_i^{\text{opt}} \sin(\omega_{\text{C}} t) + z_i^{\text{opt}}) S_z$. The overall goal of the optimization is, by repeatedly adjusting the time-varying amplitudes $\Omega_{\text{r}}(t)$ and $\Omega_{\text{i}}(t)$ of the shaped pulse, to maximize the gate fidelity under $M$ noise samples 
\begin{equation}
\label{fidelity_optimization}
\mathcal{F}_{\text{CNOT}}^{\text{opt}} = \frac{1}{M}\sum_{i=1}^M \left|\text{Tr}(U_{\text{CNOT}}^{\dagger}U_i)/4\right|^2
\end{equation}
where $U_{\text{CNOT}}$ is the ideal CNOT gate, and $\text{Tr}$ obtains the trace of a matrix. The optimization algorithm adopted here is a gradient-based algorithm, i.e., the gradient ascent pulse engineering (GRAPE) algorithm~\cite{khaneja2005optimal}. Before implementing the algorithm, for the sake of lowering the computational overhead, we let the shaped pulse take a piecewise-constant profile with the initial values randomly generated. As shown in Fig.~4(a) of the main text, the shaped pulse has 30 pieces for both real and imaginary parts (60 in total) with each piece lasting for 50 ns. Besides, it is noteworthy that only a few points are needed in the noise sampling, and (-88, 0, 88 kHz) for the static noise $Z^{\text{opt}}$ and (-76, 0, 76 kHz) for the time-varying noise $X^{\text{opt}}$ and $Y^{\text{opt}}$ are employed here for a larger noise tolerance.

\section*{Universality of our method}
Because the coupling parameters of the $^{13}$C spins are not involved in the optimization above, the optimized shaped pulse also applies to the other NV centers without very strongly coupled $^{13}$C spins (e.g., $\sim$ MHz). An alternative noise sampling with a strength of up to 150 kHz is also feasible for a better noise tolerance, which is sufficient for the NV center in the diamond with a natural abundance of $^{13}$C atoms. For the NV center with $\sim$ MHz coupled $^{13}$C spins, the $^{13}$C spins must be treated as quantum objects and included in the system Hamiltonian in the optimization process. In general, such $\sim$ MHz coupled $^{13}$C spins are often few, one or two if any, and thus will not dramatically increase the computational overhead even though they are treated as quantum objects in the optimization.

\section*{Performance of the optimized shaped pulse}
The shaped pulse is elaborately designed for realizing a high-fidelity CNOT gate with the optimization method introduced above. By numerically simulating the gate realized by the shaped pulse under the complete noise model (the five $^{13}$C spins, the static noise (20 kHz), and the time-varying noise (30 kHz)), the gate error from the $^{13}$C spin bath is estimated to be 0.074\textperthousand\ (Fig.~4(d)).

\begin{figure}
\centering
\includegraphics[width=1.0\textwidth]{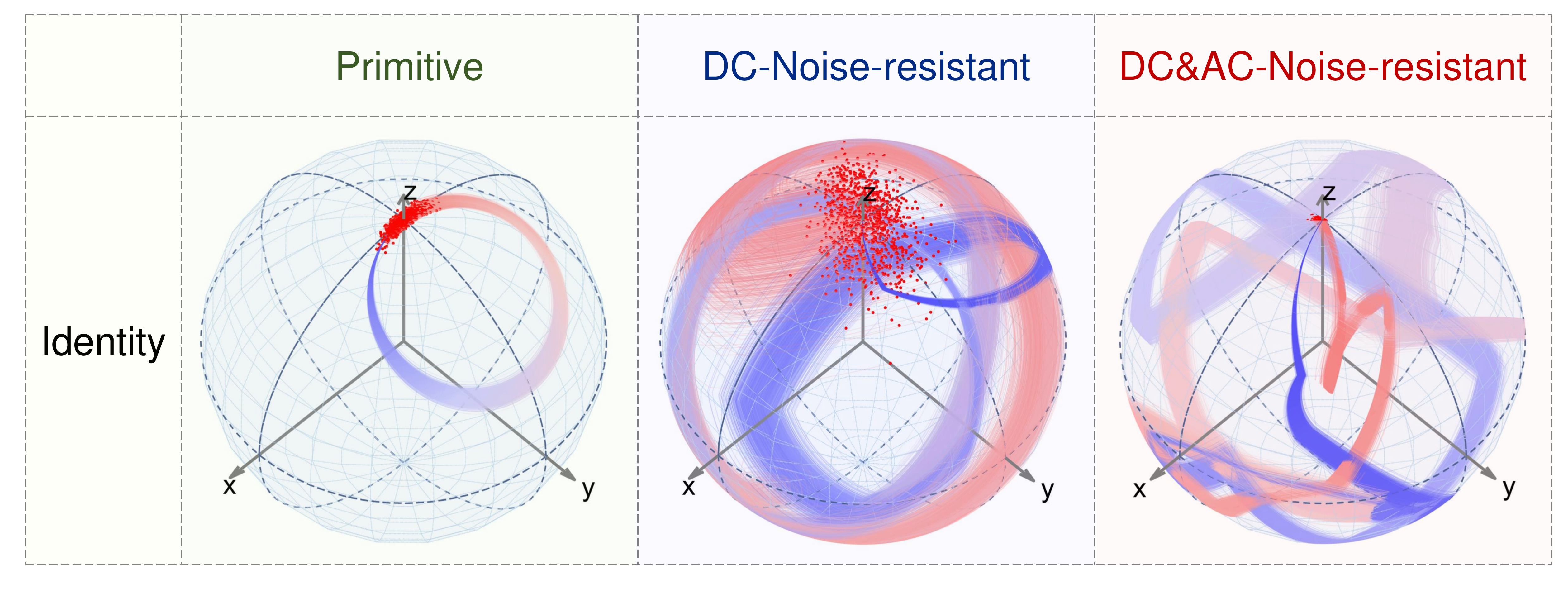}
\caption{Resistance to time-dependent noise. The state evolutions under the Identity operation of the CNOT gate for three kinds of control pulses are calculated with 1000 samples of the time-varying noise described in the main text. The results are exhibited on the Bloch sphere. See the NOT part of the CNOT gate in Fig.~2(c) of the main text.
}\label{Blochidentity}
\end{figure}

\begin{figure}
\centering
\includegraphics[width=0.75\textwidth]{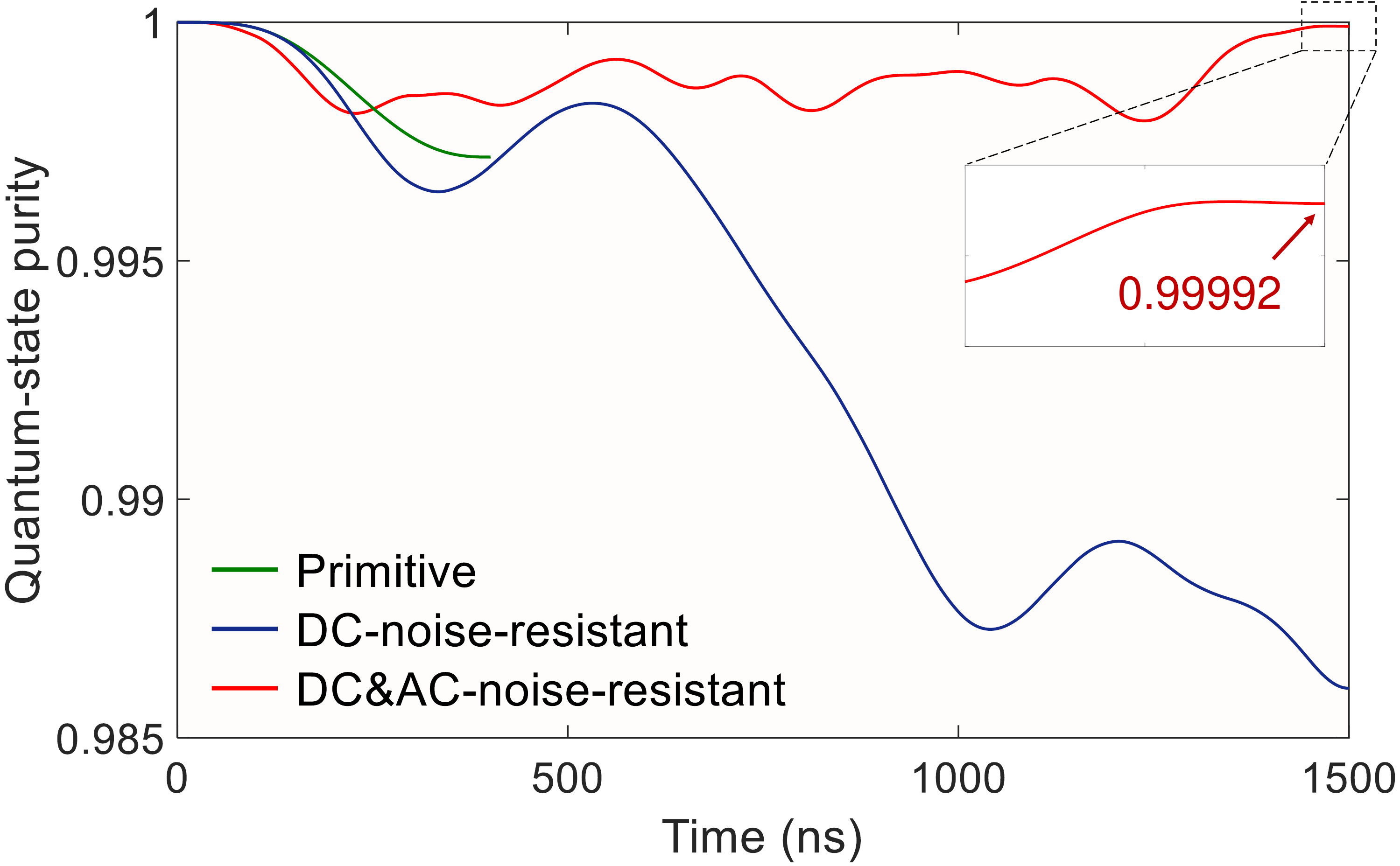}
\caption{Resistance to quantum noise. The quantum-state purity under the Identity operation of the CNOT gate is numerically calculated for comparing the abilities of three kinds of control pulses to resist the quantum noise caused by the proximal five $^{13}$C spins. See the NOT part of the CNOT gate in Fig.~2(e) of the main text.
}\label{purity_identity}
\end{figure}

Furthermore, the abilities of the shaped pulse to resist classical noise and quantum noise can be investigated separately. The performance of resisting classical noise is manifested by calculating the gate errors under time-varying noises $\eta(t) = X^\prime \cos(\omega t) + Y^\prime \sin(\omega t)$ with different frequencies $\omega$ (Fig.~2(a)). The random variables $X^\prime$ and $Y^\prime$ are zero-mean Gaussian distributions with the standard deviations given by $\sigma_x = \sigma_y = $ 70 kHz. The noise strength chosen here is a little larger than that of the fully classicized $^{13}$C spin bath (45.4 kHz) for demonstrating a better noise tolerance than demanded. The state evolutions under the shaped pulse are calculated with 1000 noise samples at the $^{13}$C Larmor frequency, as displayed in Fig.~2(a) (the NOT operation) and Fig.~\ref{Blochidentity} (the Identity operation), together with the primitive gate and the shaped pulse with only static-noise resistance for comparison. The ability to resist quantum noise is embodied by calculating the quantum-state purity under the unitary evolution with the system Hamiltonian including the nearby five $^{13}$C spins, as displayed in Fig.~2(e) (the NOT operation) and Fig.~\ref{purity_identity} (the Identity operation), together with the other two pulses for comparison.

\clearpage

\section*{Subspace randomized benchmarking}
It is an intrinsic problem for a hybrid system to directly perform two-qubit RB sequences, due to the vast difference in the quantum control for different quantum objects (here the electron spin and the nuclear spin). Fortunately, the two-qubit RB can be simplified to be performed in subspaces under some reasonable approximations, called subspace randomized benchmarking (SRB)~\cite{baldwin2020subspace}. The deviation arising therefrom is smaller than that of IRB~\cite{magesan2012efficient}, which is verified here by numerical simulation.

The gate fidelity is generally defined as the state fidelity averaged over the whole state space
\begin{equation}
\label{fidelity_gate}
\mathcal{F}_{U} = \int_{| \psi \rangle} \langle \psi | U^{\dagger} \Lambda_{U}(| \psi \rangle \langle \psi |) U | \psi \rangle d\psi
\end{equation}
where $U$ is the target gate and $\Lambda_{U}$ is the quantum operation close to $U$. The target gate here is the CNOT gate $U_{\text{CNOT}}$. The gate realized by the optimized shaped pulse is indeed a quantum operation performed upon the two-spin system, resulting from unwanted entanglement with the $^{13}$C spin bath, as shown in Fig.~2(e) and Fig.~\ref{purity_identity}. Therefore, when the $^{13}$C spin bath is considered as part of the system Hamiltonian, the whole operation will be unitary
\begin{equation}
\label{quantum_operation}
\begin{aligned}
\Lambda_{\text{CNOT}}(| \psi \rangle \langle \psi |) &= \text{Tr}_{\text{bath}}\left( U_{\text{CNOT}}^{\text{exp}}(| \psi \rangle \langle \psi | \otimes \rho_{\text{bath}}) (U_{\text{CNOT}}^{\text{exp}})^{\dagger} \right) \\
&= \frac{1}{d_{\text{bath}}} \sum_{i,j} \langle \psi_{\text{bath}}^i | U(| \psi \rangle | \psi_{\text{bath}}^j \rangle \langle \psi_{\text{bath}}^j | \langle \psi |)(U_{\text{CNOT}}^{\text{exp}})^{\dagger} | \psi_{\text{bath}}^i \rangle
\end{aligned}
\end{equation}
where $U_{\text{CNOT}}^{\text{exp}}$ is the CNOT gate realized experimentally in the whole Hilbert space. The $U_{\text{CNOT}}^{\text{exp}}$ is obtained under the interaction picture of the Larmor precession for each $^{13}$C spin so that the effect on the $^{13}$C spin bath is nearly the Identity operation. The bath state $\rho_{\text{bath}}$ is the maximum mixed state $\frac{1}{d_{\text{bath}}} \sum_{j} | \psi_{\text{bath}}^j \rangle \langle \psi_{\text{bath}}^j |$. In Eq.~(\ref{quantum_operation}), the ensemble average of the MW pulse fluctuation is ignored for simplicity considering that the error arising therefrom is rather small as shown in Fig.~4(d) of the main text. The longitudinal relaxation of the electron spin is also ignored here in that it is a depolarized channel compatible with the RB~\cite{magesan2011scalable}. Substituting Eq.~(\ref{quantum_operation}) into Eq.~(\ref{fidelity_gate}) gives the fidelity of the CNOT gate
\begin{equation}
\label{fidelity_CNOT}
\begin{aligned}
\mathcal{F}_{\text{CNOT}} &= \frac{1}{d_{\text{bath}}} \sum_{i,j} \int_{| \psi \rangle} \left| \langle \psi_{\text{bath}}^i | \langle \psi | U_{\text{CNOT}}^{\dagger} U_{\text{CNOT}}^{\text{exp}} | \psi \rangle | \psi_{\text{bath}}^j \rangle \right|^2 d\psi \\
& \approx \frac{1}{(4d_{\text{bath}})^2} \left| \text{Tr}(U_{\text{CNOT}}^{\dagger} U_{\text{CNOT}}^{\text{exp}}) \right|^2
\end{aligned}
\end{equation}
where the ideal CNOT gate $U_{\text{CNOT}}$ is also extended into the whole Hilbert space with the Identity operation on the $^{13}$C spin bath. The approximation in Eq.~(\ref{fidelity_CNOT}) holds well~\cite{pedersen2007fidelity}, which is verified by numerical calculation by taking the five $^{13}$C spin into consideration. The gate fidelity is expressed exactly as that used in the optimization with Eq.~(\ref{fidelity_optimization}) except the noise average. With Eq.~(\ref{fidelity_CNOT}), the fidelity of the CNOT can be divided as
\begin{equation}
\label{fidelity_CNOT_2}
\begin{aligned}
\mathcal{F}_{\text{CNOT}} &= \frac{1}{(4d_{\text{bath}})^2} \left| \text{Tr}(U_{\text{Ide}}^{\dagger} U_{\text{Ide}}^{\text{exp}}) + \text{Tr}(U_{\text{NOT}}^{\dagger} U_{\text{NOT}}^{\text{exp}}) \right|^2 \\
&\approx \frac{1}{4} \left( \frac{1}{2d_{\text{bath}}} \left| \text{Tr}(U_{\text{Ide}}^{\dagger} U_{\text{Ide}}^{\text{exp}}) \right| + \frac{1}{2d_{\text{bath}}} \left|\text{Tr}(U_{\text{NOT}}^{\dagger} U_{\text{NOT}}^{\text{exp}}) \right| \right)^2 \\
&= \frac{1}{4}(\sqrt{\mathcal{F}_{\text{Ide}}} + \sqrt{\mathcal{F}_{\text{NOT}}})^2
\end{aligned}
\end{equation}
Based on the equation above, the fidelity of the CNOT gate can be obtained by separately measuring the fidelities of the Identity operation and the NOT operation. The approximation in Eq.~(\ref{fidelity_CNOT_2}) is made by ignoring the error from the nuclear-spin decoherence, which gives an estimated error of $7\times10^{-6}$ in Fig.~4(d) of the main text.

\section*{Experimental implementation of randomized benchmarking}
The two-qubit system used in this work consists of the NV electron spin and the $^{14}$N nuclear spin. It can be collectively initialized into the state $|{m_{\scaleto{S}{4.5pt}}=0}\rangle |{m_{\scaleto{I}{4.5pt}}=+1}\rangle$ by a 532-nm laser pulse under a magnetic field of about 510 G, and the population of the state $|m_{\scaleto{S}{4.5pt}}=0\rangle$ can be read out by collecting fluorescence photons. The experimental sequence for implementing randomized benchmarking (RB) is plotted in Fig.~\ref{Sequence}. Apart from the main sequence for performing the RB sequence, a reference sequence is also indispensable and nearly identical to the main sequence except the absence of the RB sequence. After running the sequence, the obtained result is the ratio of the numbers of the collected photons from the main sequence, and after normalization by Rabi oscillations, the population of $|{m_{\scaleto{S}{4.5pt}}=0}\rangle$ for the fidelity measurement is obtained.

Before the fidelity measurement, the population is guaranteed to be 1 within the errorbar when the RB sequence is applied with a large detuning. The unit population is reasonable since the mixed part of the state is removed after normalization by Rabi oscillations, and does not indicate that the initial state is pure enough. The verification above eliminates varieties of unwanted disturbances affecting the results of the fidelity measurement, such as the laser power fluctuation, the drift of the NV position, the heating effect, and so on. Among them, the heating effect mainly originated from the RB sequence is especially noteworthy. The heat-expansion can drift the position of the NV center and then decrease the counting rate, which may lead to inaccurate results of the fidelity measurement. Therefore, the heating effect is measured with the results for the single-qubit RB and the IRB for the CNOT gate displayed in Fig.~\ref{Heat}(a), both of which are negligible within the errorbars. The results above also indicate that all other disturbances do not affect the fidelity measurement. The sequence for measuring the heating effect (Fig.~\ref{Heat}(b)) is similar to the sequence of Fig.~\ref{Sequence}, except that the RB sequence is off-resonant by 1 GHz and the idle stage of the reference sequence includes the duration of the RB sequence.

\begin{figure}
\centering
\includegraphics[width=1.0\textwidth]{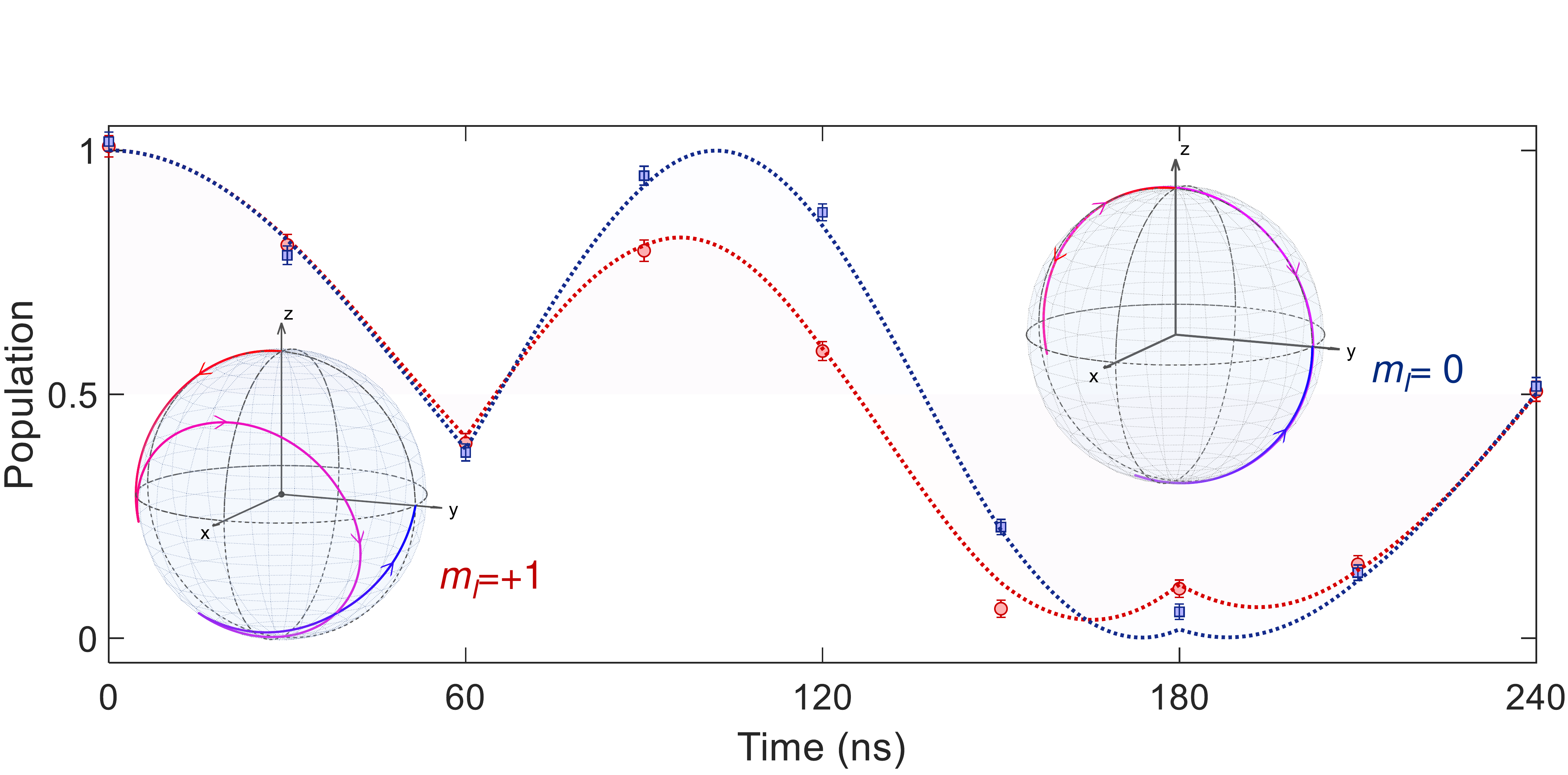}
\caption{Measured evolutions for two nuclear states under the $\pi/2$ gate. The $\pi/2$ gate is realized by applying the shaped pulse in Fig.~3(b) of the main text. The solid lines are the results of the theoretical simulation. The evolutionary trajectories on the Bloch sphere are also exhibited for a visual comparison.                                                                           
}\label{SQ_evolution}
\end{figure}

\begin{figure}
\centering
\includegraphics[width=0.7\textwidth]{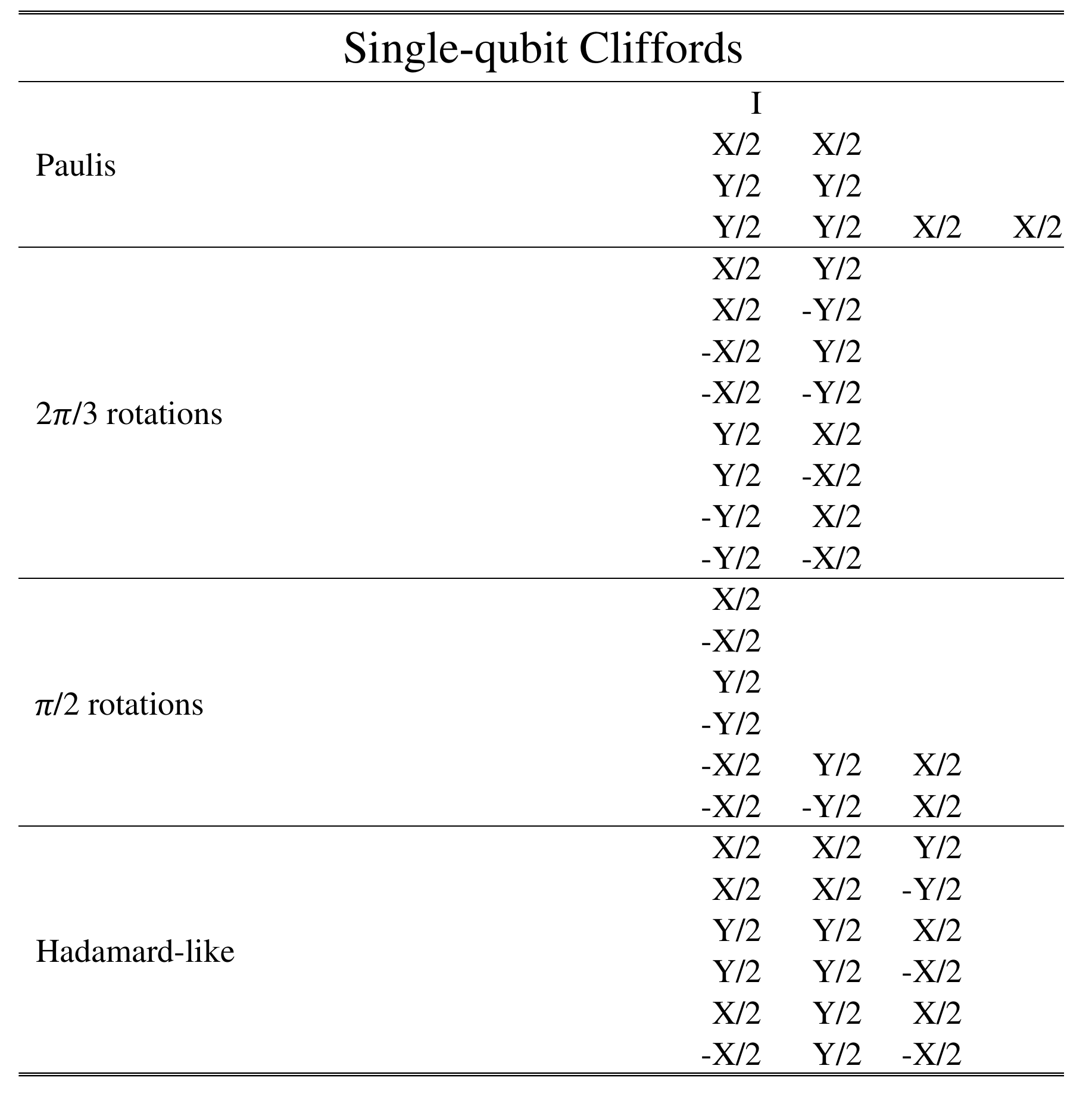}
\caption{Single-qubit Cliffords. The Single-qubit Clifford group contains 24 gates in total, and all can be generated by the $\pi/2$ gate with the shaped pulse in Fig.~3(b) of the main text.
}\label{Cliffords}
\end{figure}

\begin{figure}
\centering
\includegraphics[width=1.0\textwidth]{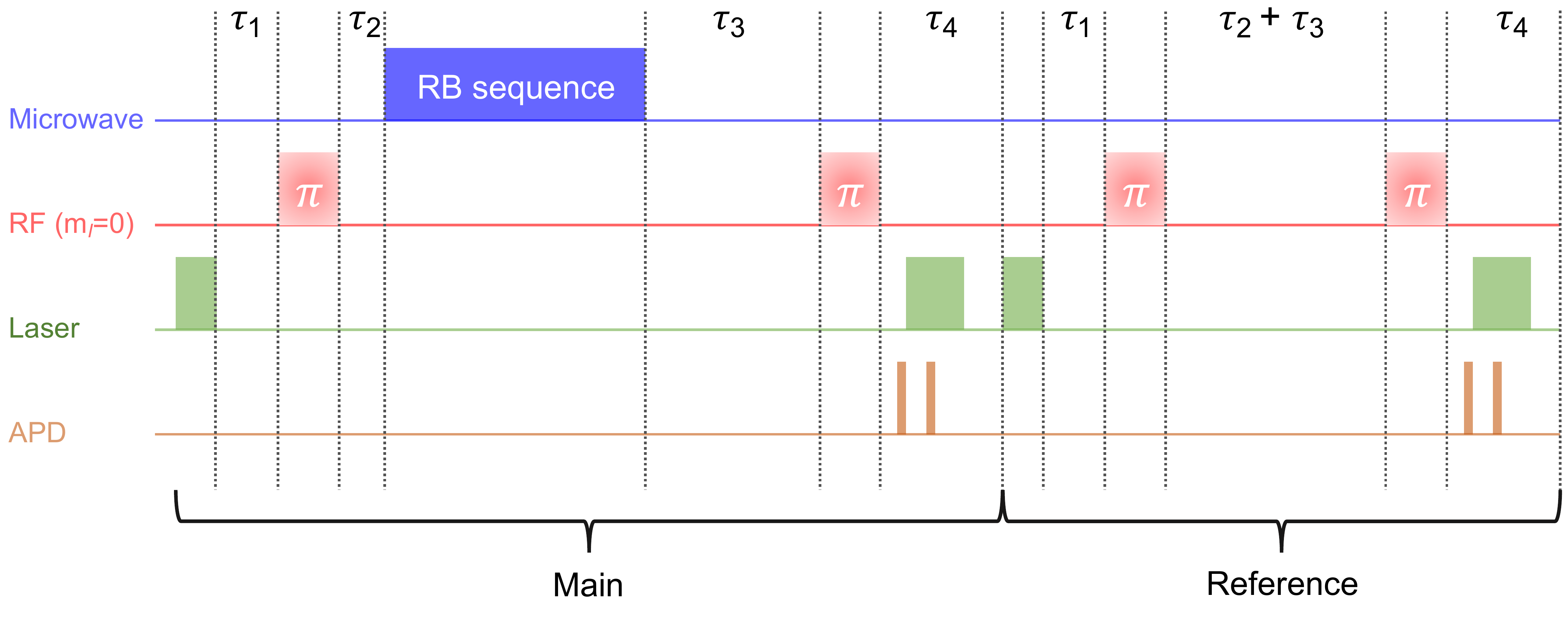}
\caption{Experimental sequence for applying randomized benchmarking. The reference sequence is nearly identical to the main sequence except the absence of the RB sequence. The RF pulses are only used for measuring the gate fidelity in the $m_{\scaleto{I}{4.5pt}} = 0$ subspace.
}\label{Sequence}
\end{figure}

\begin{figure}
\centering
\begin{overpic}[width=1.0\textwidth]{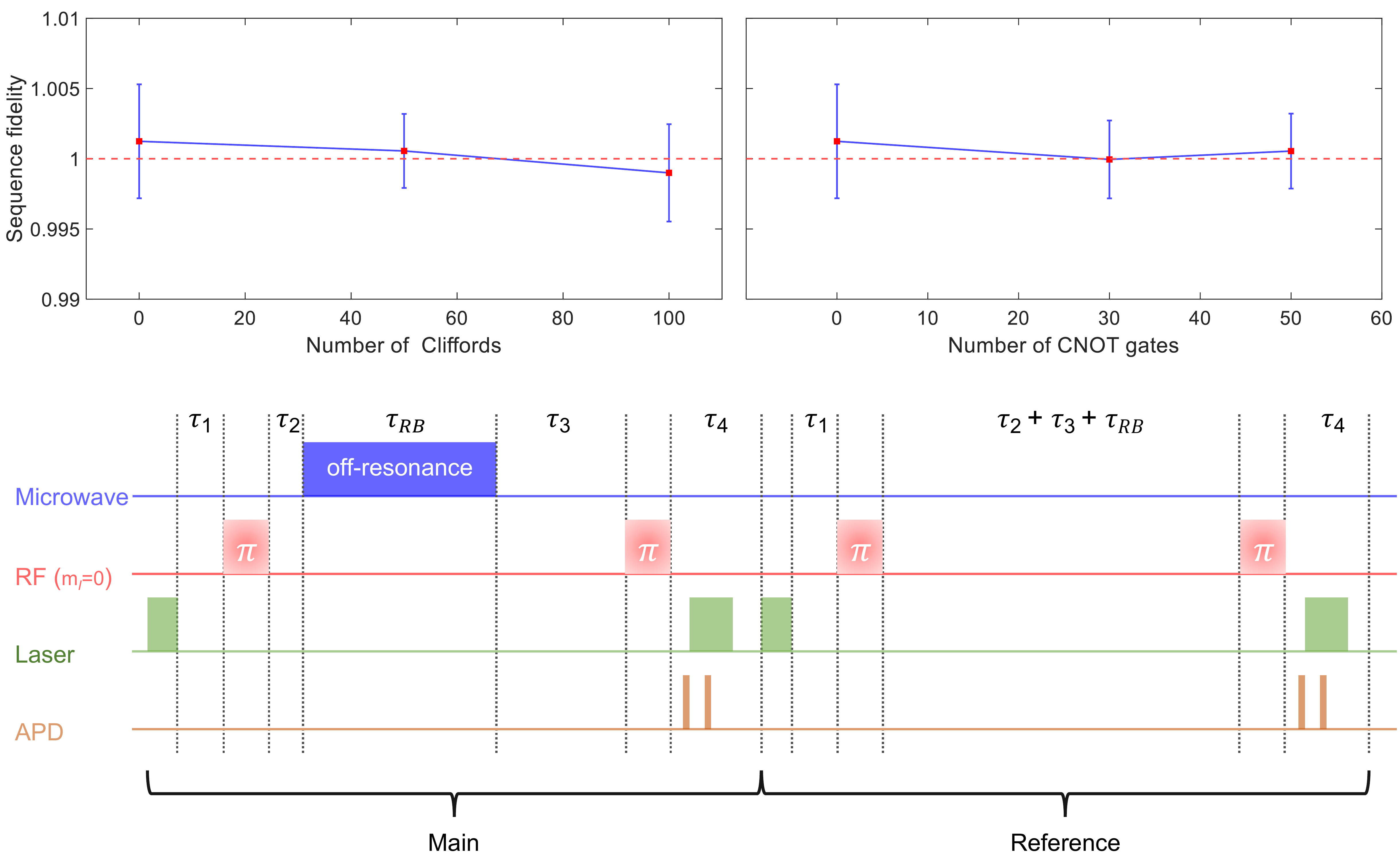}
\renewcommand{\familydefault}{\sfdefault}
	\put (0., 61.) {\figfont{(a)}}
	\put (0., 32.) {\figfont{(b)}}
\end{overpic}
\caption{Measurement of the heating effect. (a) The heating effect measured with the sequence in (b) for the single-qubit RB (left) and the IRB for the CNOT gate (right), both of which are invisible within the errorbars. (b) The sequence for measuring the heating effect mainly induced by the RB sequence and the RF pulses. Compared with Fig.~\ref{Sequence}, the RB sequence is off-resonant by 1 GHz with its duration added into the idle stage of the reference sequence. 
}\label{Heat}
\end{figure}

Before benchmarking the CNOT gate with the SRB method described above, the performance of single-qubit gates needs to be characterized first. The state evolutions in both nuclear subspaces under the $\pi/2$ gate are measured, in a good agreement with the numerical calculation as shown in Fig.~\ref{SQ_evolution}. Since the error of an individual $\pi/2$ gate is too small to be covered by the readout noise, the method of RB has to be adopted to enlarge the error by randomly concatenating many Clifford gates. With the table of Fig.~\ref{Cliffords}, all single-qubit Cliffords can be built from the basic block, i.e., the $\pi/2$ gate realized by the shaped pulse in Fig.~3(b) of the main text. Each Clifford gate contains $13/6$ gates on the average. The $\pi/2$ gate here is a true single-qubit gate that implements the same operation in both nuclear subspaces, rather than only working in the $m_{\scaleto{I}{4.5pt}}=+1$ as done by the previous work~\cite{rong2015experimental}. When the RB sequence is carried out in the $m_{\scaleto{I}{4.5pt}}{=+1}$ subspace, a RF $\pi$ pulse for driving the transition $|{m_{\scaleto{S}{4.5pt}}=0}\rangle |{m_{\scaleto{I}{4.5pt}}=+1}\rangle \leftrightarrow |{m_{\scaleto{S}{4.5pt}}=0}\rangle |{m_{\scaleto{I}{4.5pt}}=0}\rangle$ is applied before and after the RB sequence to prepare and readout the population of the state $|{m_{\scaleto{S}{4.5pt}}=0}\rangle |{m_{\scaleto{I}{4.5pt}}=+1}\rangle$. The single-qubit RB and the SRB for evaluating the CNOT gate are executed in both nuclear subspaces with the results described in the main text.

\clearpage

\section*{Gate error analysis}
Although the error from the reservoir is largely suppressed by the DEC effect of the shaped pulse, the errors from the longitudinal relaxation of the electron spin and the waveform distortion of the shaped pulse become more significant, due to a more complicated pulse composition and a longer gate time compared with the primitive gate. Apart from the two most important factors, all error sources are detailedly analyzed here, except the error from $^{13}$C spin bath discussed above, with the results summarized in the table of Fig.~4(d).

\begin{figure}
\centering
\includegraphics[width=0.7\textwidth]{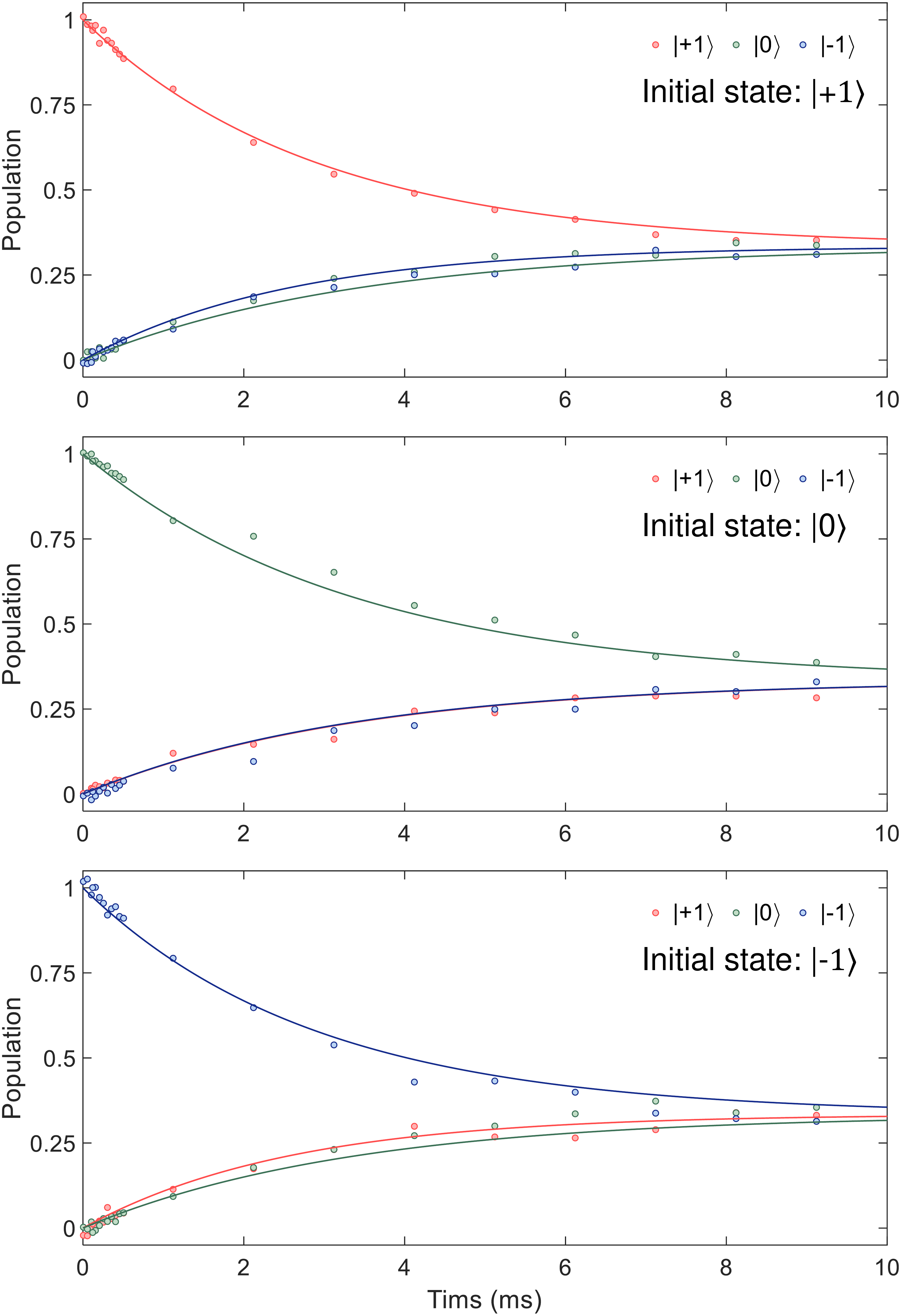}
\caption{Three-level longitudinal relaxation. The relaxation behaviors of the NV electron spin ($S=1$) are measured with different initial states marked in each graph. The red, green, and blue dots stand for the states $|+1\rangle$, $|0\rangle$, and $|-1\rangle$, respectively. The solid lines are plotted by simultaneously fitting the nine groups of data points above.
}\label{t1}
\end{figure}

First of all, longitudinal relaxation behaviors of the NV spin triplet are measured for estimating the error arising therefrom. When the initial state is prepared at the eigenstates $m_{\scaleto{S}{4.5pt}}=0,\pm1$, the process of longitudinal relaxation can be described by a rate equation
\begin{equation}
\label{rate_equation}
\dot{\rho}(t) = \Gamma \rho(t)
\end{equation}
where the state $\rho(t)$ is not a density matrix but a column vector containing the populations of $m_{\scaleto{S}{4.5pt}}=+1, 0, -1$. $\Gamma$ is the decay matrix given by
\begin{equation}
\label{relaxation}
\Gamma = 
\begin{pmatrix}
-\gamma_{+1}-\gamma_{2} & \gamma_{+1} & \gamma_{2}\\
\gamma_{+1} & -\gamma_{+1}-\gamma_{-1} & \gamma_{-1}\\
\gamma_{2} & \gamma_{-1} & -\gamma_{-1}-\gamma_{2}\\
\end{pmatrix}
\end{equation}
where $\gamma_{\pm1}$ is the decay rate between $m_{\scaleto{S}{4.5pt}}=\pm1$ and $m_{\scaleto{S}{4.5pt}}=0$, and $\gamma_{2}$ is the rate between $m_{\scaleto{S}{4.5pt}}=+1$ and $m_{\scaleto{S}{4.5pt}}=-1$. Here, $\gamma_{+1}$ and $\gamma_{-1}$ are not assumed to be identical for generality. By solving the corresponding secular equation, the eigenvalues and eigenvectors of the decay matrix are obtained
\begin{gather}
\lambda_1 = 0, \lambda_2 = -(\gamma_{+1}+\gamma_{-1}+\gamma_{2}) + x_0, \lambda_3 = -(\gamma_{+1}+\gamma_{-1}+\gamma_{2}) - x_0 \label{eigenvalues} \\
(X_1, X_2, X_3) =
\begin{pmatrix}
1 & \gamma_{+1}\gamma_{-1}+\gamma_{2}x_0-\gamma_{2}^2 & \gamma_{+1}\gamma_{-1}-\gamma_{2}x_0-\gamma_{2}^2\\
1& \gamma_{+1}\gamma_{2}+\gamma_{-1}x_0-\gamma_{-1}^2 & \gamma_{+1}\gamma_{2}-\gamma_{-1}x_0-\gamma_{-1}^2\\
1 & (\gamma_{2}-x_0)(\gamma_{-1}-x_0)-\gamma_{+1}^2 & (\gamma_{2}+x_0)(\gamma_{-1}+x_0)-\gamma_{+1}^2 \label{eigenvectors} \\
\end{pmatrix}
\end{gather}
where $x_0=\sqrt{\gamma_{+1}^2+\gamma_{-1}^2+\gamma_{2}^2-\gamma_{+1}\gamma_{-1}-\gamma_{+1}\gamma_{2}-\gamma_{-1}\gamma_{2}}$. The eigenvectors in Eq.~(\ref{eigenvectors}) are not normalized. Accordingly, the state evolutions with different initial states $m_{\scaleto{S}{4.5pt}}=+1$, $0$, and $-1$ are given by
\begin{eqnarray}
\label{population_evolution}
\nonumber \rho_{+1}(t) &=& \frac{X_1}{3} + \frac{\gamma_{+1}\gamma_{-1}+\gamma_{2}x_0-\gamma_{2}^2}{||X_2||^2} e^{\lambda_3 t} X_2 + \frac{\gamma_{+1}\gamma_{-1}-\gamma_{2}x_0-\gamma_{2}^2}{||X_3||^2} e^{\lambda_2 t} X_3\\
\rho_{0}(t) &=& \frac{X_1}{3} + \frac{\gamma_{+1}\gamma_{2}+\gamma_{-1}x_0-\gamma_{-1}^2}{||X_2||^2} e^{\lambda_2 t} X_2 + \frac{\gamma_{+1}\gamma_{2}-\gamma_{-1}x_0-\gamma_{-1}^2}{||X_3||^2} e^{\lambda_3 t} X_3\\
\nonumber \rho_{-1}(t) &=& \frac{X_1}{3} + \frac{(\gamma_{2}-x_0)(\gamma_{-1}-x_0)-\gamma_{+1}^2}{||X_2||^2} e^{\lambda_2 t} X_2 + \frac{(\gamma_{2}+x_0)(\gamma_{-1}+x_0)-\gamma_{+1}^2}{||X_3||^3} e^{\lambda_3 t} X_3
\end{eqnarray}
where $\rho_{+1}(t)$, $\rho_{0}(t)$, and $\rho_{-1}(t)$ correspond to the evolutions with initial states $m_{\scaleto{S}{4.5pt}}=+1$, $0$, and $-1$, respectively. Based on Eq.~(\ref{population_evolution}), nine experiments are performed with all three states prepared and readout after a waiting time, with the results shown in Fig.~\ref{t1}. By simultaneously fitting all data points with the nine functions from Eq.~(\ref{population_evolution}), the decay matrix is determined to be
\begin{equation}
\label{relaxation_results}
\Gamma = 
\begin{pmatrix}
-229(6) & 98(4) & 130(5)\\
98(4) & -198(6) & 100(4)\\
130(5) & 100(4) & -230(6)\\
\end{pmatrix}
\end{equation}
where the unit of each value in the matrix is s$^{-1}$, and all values in parentheses stand for one standard deviation. With the equation $((\gamma_{+1}+\gamma_2)/2+\gamma_{-1})T_{\text{CNOT}}$, the error from the longitudinal relaxation is estimated to be 0.32\textperthousand.

\begin{figure}
\centering
\begin{overpic}[width=0.48\textwidth]{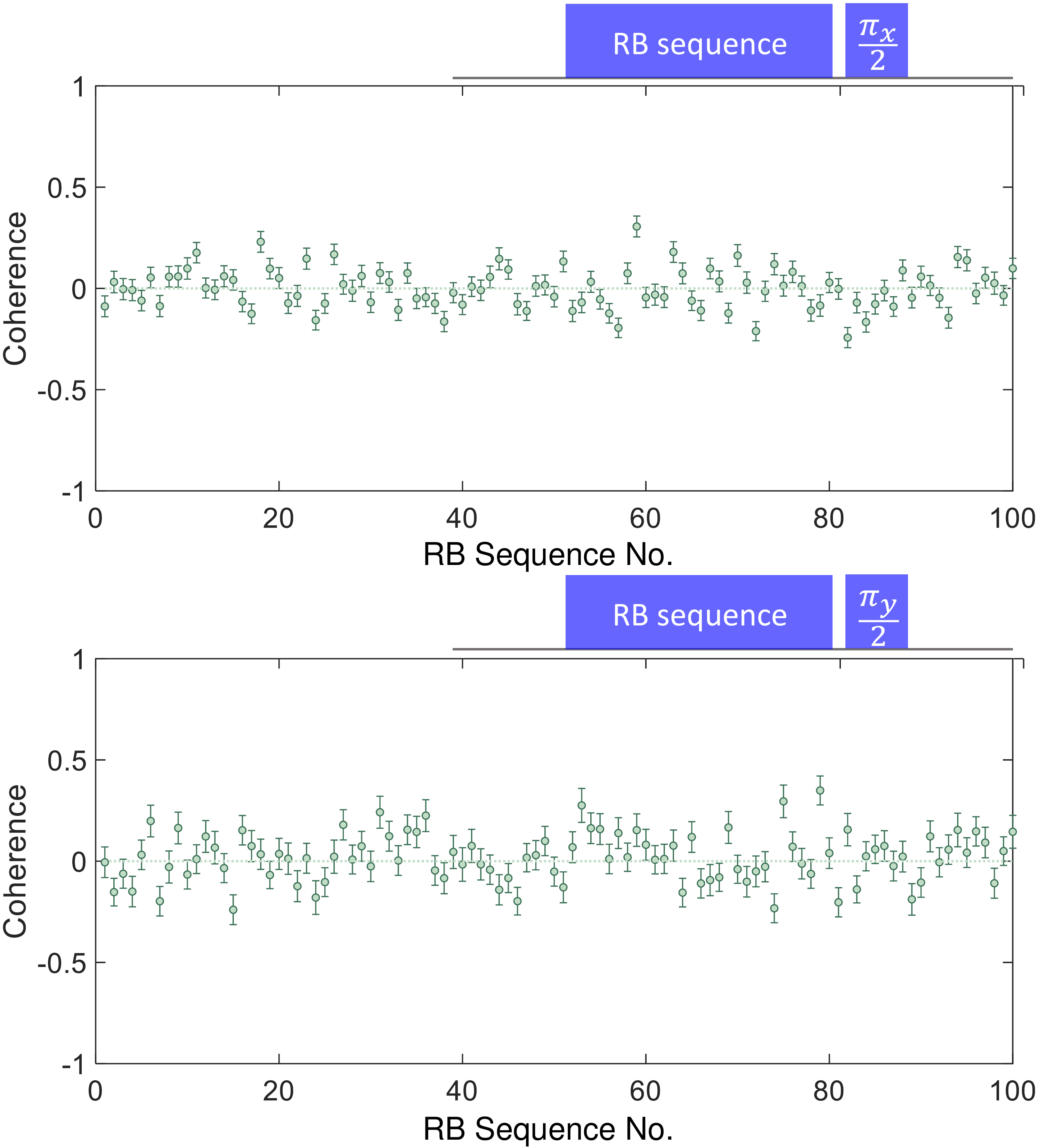}
\end{overpic}
\begin{overpic}[width=0.48\textwidth]{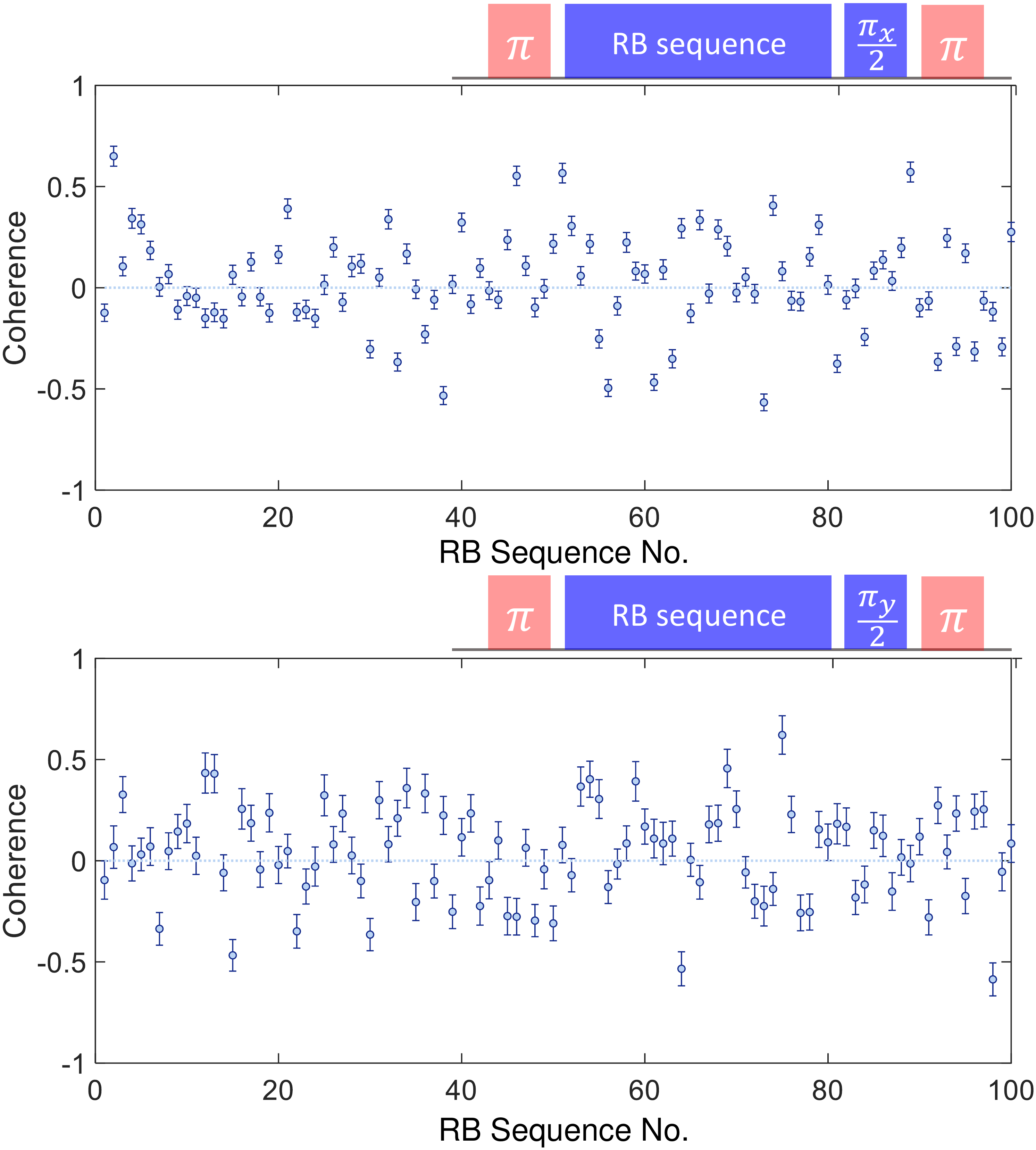}
\end{overpic}
\caption{State tomography after single-qubit RB. The state tomography after the 100-Clifford RB sequences in both $m_{\scaleto{I}{4.5pt}} = +1$ (left) and $m_{\scaleto{I}{4.5pt}} = 0$ (right) subspaces is performed for evaluating the error from the waveform distortion of the shaped pulse. The results of the $\pi_x/2$ gate (upper) an the $\pi_y/2$ gate (lower) together with the corresponding RB result (no gate) constitute a full single-qubit tomography.
}\label{Tomo_RB}
\end{figure}

\begin{figure}
\centering
\begin{overpic}[width=0.48\textwidth]{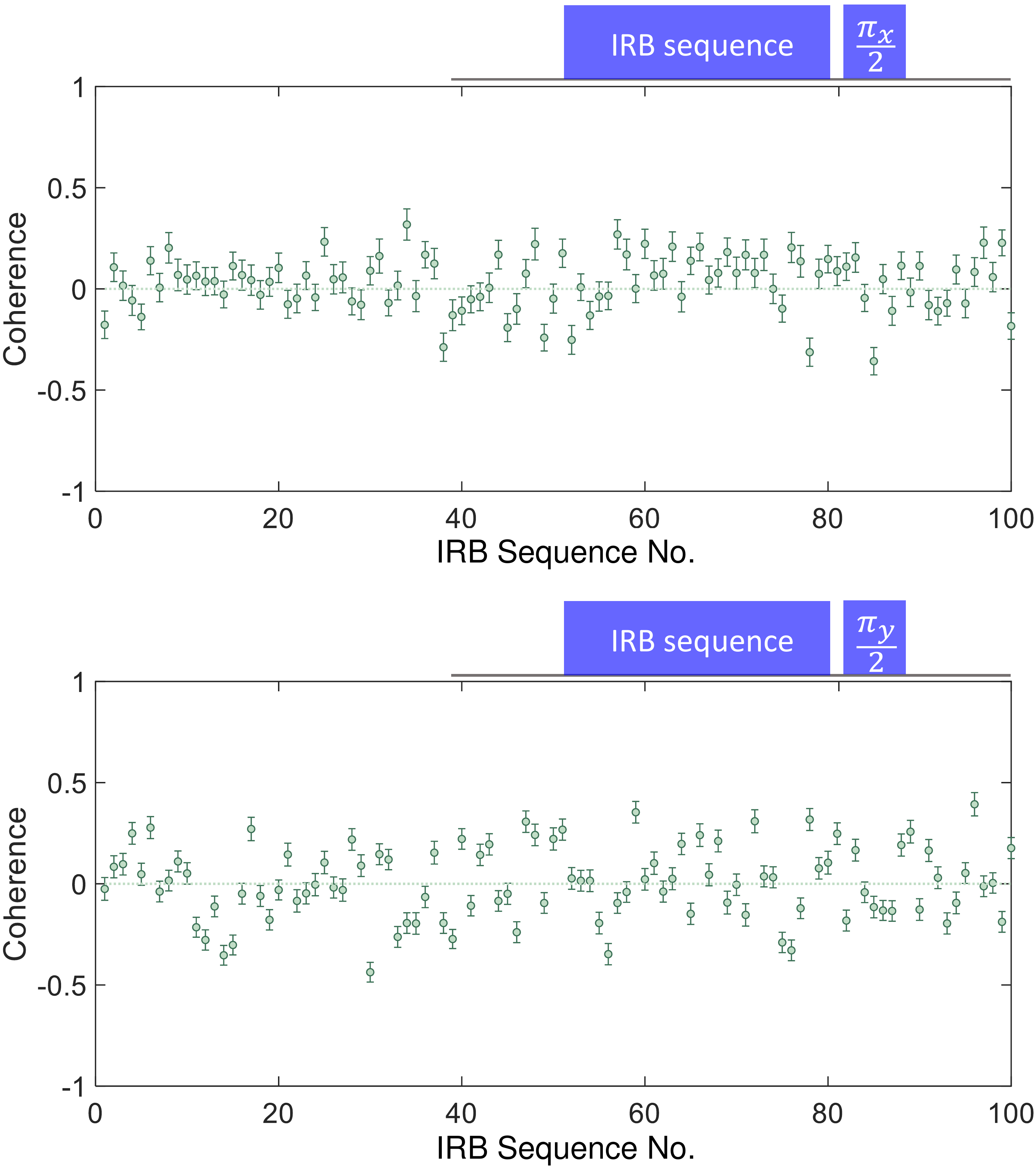}
\end{overpic}
\begin{overpic}[width=0.48\textwidth]{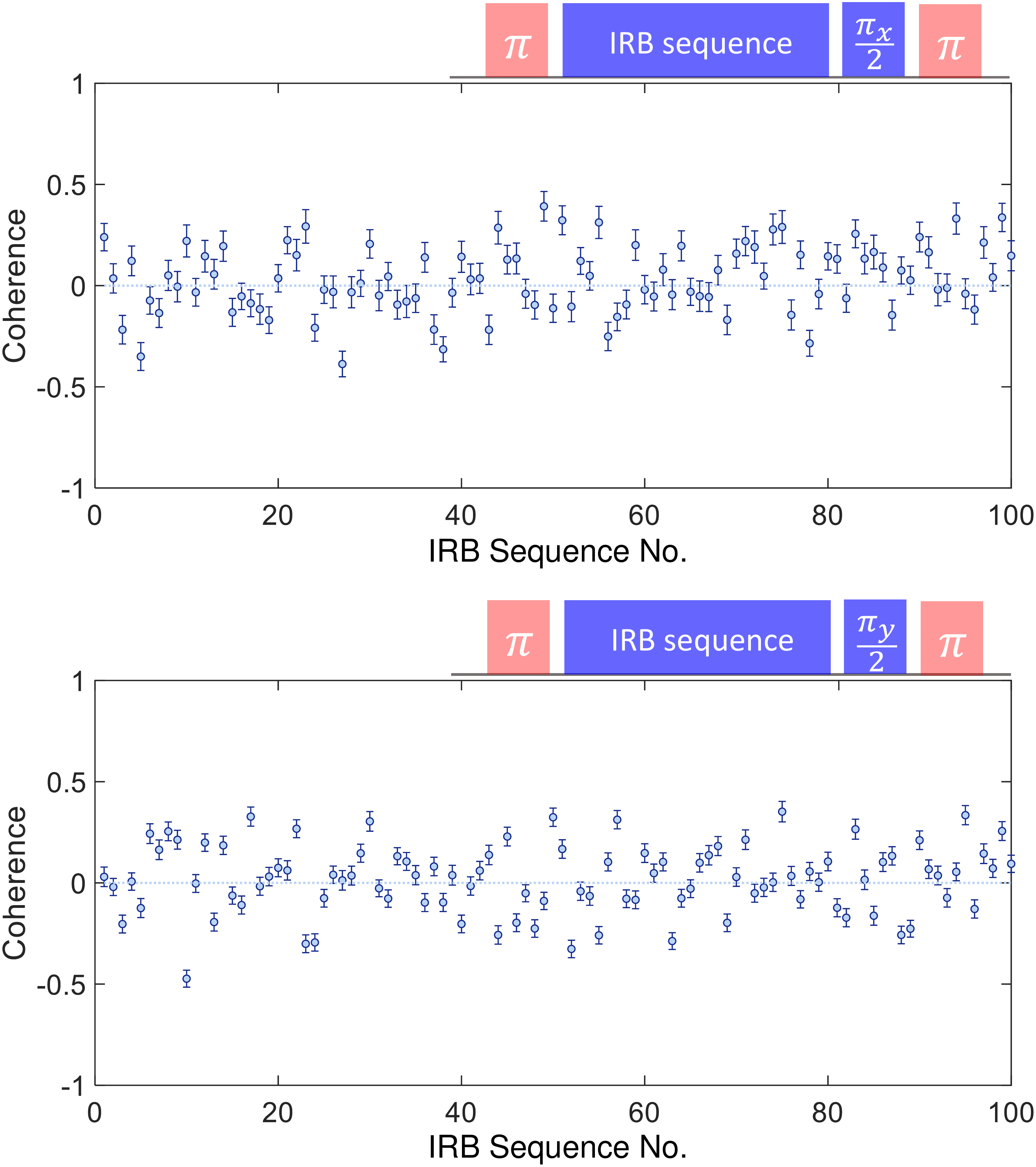}
\end{overpic}
\caption{State tomography after CNOT-gate IRB. The state tomography after the IRB sequences with 50 CNOT gates included is performed in both $m_{\scaleto{I}{4.5pt}} = +1$ (left) and $m_{\scaleto{I}{4.5pt}} = 0$ (right) subspaces for evaluating the error from the waveform distortion of the shaped pulse. The results of the $\pi_x/2$ gate (upper) an the $\pi_y/2$ gate (lower) together with the corresponding IRB result (no gate) constitute a full single-qubit tomography.
}\label{Tomo_IRB}
\end{figure}

Compared with the primitive gate, the gate realized by the shaped pulse in realistic settings, inevitably suffers more from the problem of waveform distortion, owing to the complicated profile optimized for the effect of DEC. Unlike the errors from various fluctuations, the waveform distortion gives rise to a deterministic error, namely, a unitary error. Thus, the error from the waveform distortion can be estimated by constructing the density matrix after implementing each RB and IRB sequence. The population of the state after each sequence has been provided by the RB and IRB results, while the real part and the imaginary part of the coherence are measured by applying a $\pi_y/2$ pulse and a $\pi_x/2$ pulse respectively before reading out the state, with the results shown in Fig.~\ref{Tomo_RB}-\ref{Tomo_IRB}. With the density matrix, the quantum-state purity is calculated for the state after implementing each sequence, and represents an estimated fidelity without the waveform distortion. The fidelity difference with and without the waveform distortion is the error arising therefrom. The error on the CNOT gate from the waveform distortion is estimated to be 0.21\textperthousand\ from the IRB results after subtracting that on single-qubit Cliffords from the RB results.

\begin{figure}
\centering
\includegraphics[width=0.75\textwidth]{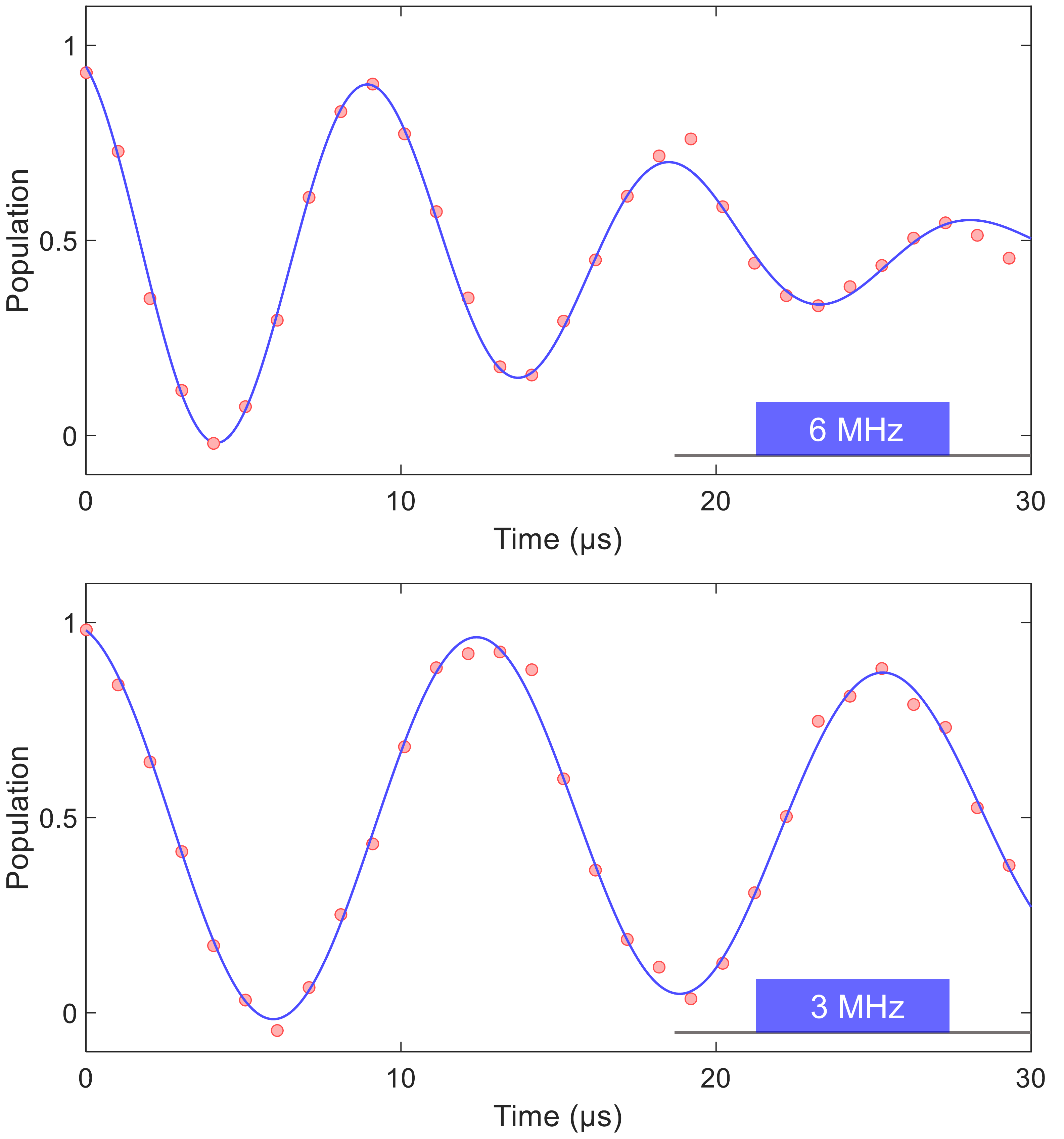}
\caption{Measured instability of microwave amplitude. Undersampling measurements of the Rabi oscillations with two amplitudes (6 MHz and 3 MHz) are used for evaluating the instability of the microwave amplitude. The data in each graph are accumulated by running the experiment for about 5 hours, and the solid lines are the fitting curves with the function $a\cos(2\pi(\delta f)t+\phi_0)e^{-(t/T_2^\prime)^2}+b$.
}\label{Stability}
\end{figure}

Gate errors from other sources are small enough. First, we analyze the error from the instability of the MW amplitude. As shown in Fig.~\ref{Stability}, Rabi oscillations with long durations are measured in an undersampling manner for the amplitudes 6 MHz and 3 MHz. The decay behavior is close to $e^{-(t/T_2^{\prime})^2}$, different from that in the work~\cite{rong2015experimental}. Therefore, the normalized amplitude with respect to the absolute value (6 MHz and 3 MHz), is supposed to comply with a Gaussian distribution, and the instabilities are fitted to be 0.018(2) for 6 MHz and 0.015(2) for 3 MHz. Here we use the value of 0.018 in the simulation to take up an estimated error of 0.030\textperthousand. Second, the gate error is caused by the approximations used for deriving the system Hamiltonian (\ref{system_interaction}), including the rotation wave approximation (RWA) and the reduction into the four-level structure. The latter gives rise to the population leakage into the third level $m_{\scaleto{S}{4.5pt}}=+1$. The arising error is given by 0.020\textperthousand\  by calculating the gate realized by the shaped pulse under the natural Hamiltonian (\ref{system}). Third, the $^{14}$N-nuclear-spin decoherence, mainly caused by the applied pulse not by the $^{13}$C spin bath, will disturb the superposition between two nuclear subspaces. The error from the nuclear decoherence is estimated to be $7\times10^{-6}$, which justifies the assumption for the method of SRB. The errors from other sources are all below $10^{-6}$, including the inaccuracy of the coupling parameter used for the optimization of the shaped pulse (2159.88 kHz),  the nuclear decoherence from the $^{13}$C spin bath, the sideband effect from the $z$ component of the MW pulse, the manipulation of the MW pulse on the $^{14}$N nuclear spin, and so on.

\clearpage

\bibliography{references.bib}
\clearpage

\clearpage

